\documentclass[12pt]{article}

\usepackage{graphicx,amsmath,amssymb,color,bm}
\begin{document}

\centerline {\Large\textbf {Stacking-configuration-enriched essential properties}}
\centerline {\Large\textbf {in bilayer silicenes}}

\centerline {\Large \textbf {}}\vskip0.6 truecm

\centerline{Hsin-yi Liu$^{1,2}$, Shih-Yang Lin$^{3}$, and Jhao-ying Wu$^{4,*}$}

\centerline{$^{1}$Center for Micro/Nano Science and Technology, National Cheng Kung University, Tainan, Taiwan}
\centerline{$^{2}$Department of Physics/QTC/Hi-GEM, National Cheng Kung University, Tainan, Taiwan}
\centerline{$^{3}$Department of Physics, National Chung Cheng University, Chiayi, Taiwan}
\centerline{$^{4}$Center of General Studies, National Kaohsiung University of Science and Tachnology, Kaohsiung, Taiwan}

\begin{abstract}


The geometric, electronic and magnetic properties of silicene-related systems present the diversified phenomena through the first-principles calculations. The critical factors, the group-IV monoelements,  buckled/planar structures, stacking configurations, layer numbers, and van der Waals interactions of bilayer composites are taken into account simultaneously. The developed theoretical framework is responsible for the concise physical and chemical pictures. The delicate evaluations and analyses are conducted on the optimal lattices, the atom- $\&$ spin-dominated energy bands, the atom-, orbital- $\&$ spin-projected vanHove singularities, and the magnetic moments. Most importantly, they achieve the decisive mechanisms, the buckled/planar honeycomb lattices, the multi-/single-orbital hybridizations, and the significant/negligible spin-orbital couplings. Furthermore, we investigate the stacking-configuration-induced dramatic transformations of the essential properties by the relative shift in bilayer graphene and silicene. The lattice constant, interlayer distance, the buckle height, and the total energy essentially depend on the magnitude and direction of the relative shift: AA $\rightarrow$ AB $\rightarrow$ AA$^{\prime}$ $\rightarrow$ AA. Apparently, sliding bilayer systems are quite different between silicene and graphene in terms of electronic properties, strongly depending on the buckled/planar honeycomb lattices, the multi-/single-orbital hybridizations, the dominant/observable interlayer hopping integrals, and the significant/negligible spin interactions. The predicted results can account for the up-to-date experimental measurements.

\end{abstract}

\vskip0.6 truecm

\noindent

\newpage

\section{Introduction}

The emergent two-dimensional materials have attracted a lot of experimental and theoretical researches since of the first discovery of few-layer graphene systems by the mechanical exfoliation in 2004 \cite{SCI306;666}. Up to now, the mono-element group-IV and group-V systems cover silicene \cite{NC7;10657}, germanene \cite{SR6;20714}, tinene \cite{NM14;1020}, phosphorene \cite{ACSNANO8;4033}, antimonene \cite{NANOLETT18;2133, AM29;1605299}, and bismuthene \cite{AM29;1605299,NANOSCALE10;21106}. Such layered systems are very suitable for studying the diverse physical, chemical and material phenomena, mainly owing to the various Hamiltonians with the rich and unique intrinsic geometric symmetries and atomic interactions. Furthermore, they are expected to have the highly potential applications, such as, nano-electronics \cite{RSCA8;11799, AIP9;025120, JJAP54;040102, RSCA3;26153, SR5;09075, NC7;13352, NRL10;254, JMCC7;9195, ANGC58;134}, optoelectronics \cite{NANOSCALE10;21106, NC5;3389, AOM7;1800441, NP4;611, PRX4;021029, JMCC4;5434, LPR12;1700221, PCCP19;3660}, and energy storage \cite{CSR47;3189, EES11;772, CSR47;6370, PCCP21;4276, NANOLETT15;2510, NANOSCALE8;7272, CC55;3983, AM15;18, NANOSCALE9;13384, JMCA7;3238, AEM8;1702606, ANGC58;1574, NC9;1813, JMCC6;7976}. Very interesting, how to thoroughly solve and comprehend the essential properties in the atom-dependent layered materials becomes one of the main-stream topics. Apparently, the critical factors, being based on the important pictures, mainly originate from the planar/non-planar lattices, stacking configurations, layer numbers, intralayer $\&$ interlayer multi-/single-orbital hybridizations, and intrinsic magnetisms/spin-orbital couplings. The focuses of this work are the diversified fundamental properties in bilayer silicene systems through the relative shifts between two layers, in which the first-principles calculations, with the spin configurations, are available in the thorough investigations.

The group-IV layered systems, which consist of C \cite{NANOLETT18;2133, DRM88;151}, Si \cite{NC7;10657, PCCP16;304}, Ge \cite{SR6;20714,SR7;40600}, Sn \cite{NM14;1020, NP14;344} and Pb \cite{SS257;259, JPCA115;7096} atoms, have been successfully produced in experimental laboratories since the first discovery of few-layer graphene systems by the mechanical exfoliation in 2004 \cite{SCI306;666}. Only the carbon-related 2D materials are synthesized by the various growth methods, while the others are produced under the molecular beam epitaxial growth and the mechanical method. By using STM [or combination with LEED/RHEED], the graphene-like honeycomb lattices are clearly revealed in group-IV 2D systems, such as, 2D C-, Si- Ge-, Sn-, and Pb-adlayer, with the hexagonal symmetries, respectively, on SiO$_{2}$/Cu/Bi$_2$Se$_3$ \cite{SR7;43756, NANOLETT10;4328, SR9;4791},Ag(111), Ir(111) and ZrB$_{2}$(0001) \cite{PCCP10;1039, SS608;297, NANOLETT13;685, 2DM4;021015, JCP144;134703, APL110;041601}, Pt(111)/Al(111/Au(111) \cite{AM26;4820, PCCP19;18580, JPCC123;12910}, Bi$_2$Te$_3$[111] \cite{JAC611;313, ASS353;232}, and Au(111)/AlO $\&$ Si(111) \cite{SS257;259, JPCA115;7096}. Among of them, graphene systems exhibit the only planar structure in the absence of two-sublattice buckling, mainly owing to the dominating sp$^2$ bonding. The other group-IV materials are thoroughly examined to possess the buckled structures arising from the height difference between A and B sublattices \cite{NC4;1500, JPCC119;11896}.

The up-to-date data show that the width and buckling of each layer are 0 $\AA$, 0.46 $\AA$, 0.68 $\AA$, 0.84 $\AA$, and 0.89 $\AA$, respectively. The buckling degree grows in the increase of atomic number, clearly indicating the enhanced strength of sp$^3$ bonding. As for the experimental examinations on the band properties near the Fermi level, the semiconducting and semimetallic behaviors in layered graphenes are confirmed by the STS and ARPES measurements. The latter are also utilized to examine the presence or absence of Dirac-cone-like energy bands in other group-IV systems, clearly illustrating the critical roles of the adlayer-substrate chemical bondings. Obviously, the semiconducting and metallic adlayers have become one of the main-stream 2D materials, in which a plenty of essential properties are worthy of the systematic investigations. In addition, few-layer graphene systems can exhibit the rich and unique physical, chemical, and material phenomena, e.g., the best mechanical property \cite{NP6;30, JPCC116;8271}, the stacking- $\&$ layer-number-diversified band properties, the diverse optical/magneto-optical spectra \cite{NP4;611, JPCM25;085508, PRB98;195442, APL107;263101, PRB84;113412, PRB94;165428, PRB84;125455, PRB95;075422, SR8;11070, PCCP17;26008, PCCP18;17597, NJP15;015010, APE9;065103, RSCA4;63779, APL98;261920}, the unusual quantum Hall conductivities \cite{NP7;948, NAT438;201, NP7;953}, and the various single- and many-particle Coulomb excitations \cite{PRB74;085406, PRB62;8508}.

Also, the group-IV 2D pristine materials have attracted a lot of theoretical studies. In general, there exist thee methods, the first-principles method, the tight-binding model and the effective-mass approximation, in exploring the electronic properties. They are available in studying energy bands, subenvelops functions and density of states. Only the first one is suitable in investigating the optimal geometric structures and the orbital hybridizations in chemical bonds. According to the up-to-date theoretical predictions, monolayer group-IV systems are direct- or indirect-gap semiconductors, in which energy gap grows from zero in the increase of atomic number. Also noticed that a single-layer graphene is a gapless semiconductor because of the vanishing density of states at the Fermi level. As to the low-lying energy bands, C-, Si- and Ge-related systems are predicted to be dominated by the $\pi$ bondings associated with the ${p_z}$ orbitals, while Sn- and Pb-dependent ones present the significant sp$^3$-orbital hybridizations. Such results are mainly determined by the buckling degree, the distribution ranges of the outer orbitals, and the spin-orbital couplings. The direct combination of the intrinsic critical factors, magnetic field and gate voltages will create the diversified essential properties, e.g., the magneto-electronic and optical properties  in few-layer graphene \cite{PCCP17;26008, PCCP18;17597, NJP15;015010, APE9;065103, RSCA4;63779, APL98;261920}, silicene \cite{PRX4;021029,JAP116;024303, SR9;36547, PCCP19;2148, PRB97;125416}, germanene \cite{JMCC4;5434} and tinene \cite{SR7;1849}. In addition, the generalized tight-binding model has been well developed to achieve a full understanding of the diverse magnetic quantization phenomena for layered graphene systems under the various dimensions \cite{PRL98;206802, PRL110;146802, NP3;36, ACSNANO6;6930, PRB73;045124, PRL100;037601}, geometric symmetries \cite{DRM88;151, JPCC119;11896, NR1;497, SIAM56;499, RMP81;109}, stacking configurations \cite{SCI313;951, NM12;887892, NANOLETT9;2654}, layer numbers \cite{ PRL98;206802, PRB88;155439}, and external fields \cite{SR4;7509}, However, the effective-mass method, covering the perturbation concept, cannot deal with the magnetically quantized states from multi-pair, unusual, or complicated valence and conduction bands \cite{PCCP17;26008, PRB73;144427}, e.g., the Landau levels in Sn- and Pb-few-layer systems.


Within the theoretical prediction, there exist a lot of meta-stable stacking configurations in few-layer graphene \cite{JPCC119;3818, JPCC119;10623}. The significant differences between the sliding bilayer silicene and graphene systems \cite{JPCC119;3818, JPCC119;10623} are worthy of the systematic investigations, mainly owing to the intrinsic geometric structures, intralayer $\&$ interlayer hopping integrals, chemical bondings and spin configurations. The above factors could lead to the diversified electronic properties (finite-gap semiconductors, zero-gap semiconductors, semimetals and metals) and magnetic configurations (non-magnetism, ferromagnetism, and anti-ferromagnetism).


The evaluated results of sliding bilayer silicene cover the significant dependencies on the relative shift between the upper and lower layers for the total energy, lattice constant, inter-sublattice distances, buckling angles, electronic energy spectra, energy gaps, spatial charge densities, density of states, magnetic moments, and spin density distributions.Furthermore, the layer-, sublattice-, orbital- and spin-decomposed physical quantities are very useful in getting the critical orbital hybridizations and the magnetic configurations. Specifically, the zero-gap semiconductors, finite-gap ones, and semimetals will clearly be identified from the specific density of states at the Fermi level. Specifically, the valley structures, which are initiated from the various band-edge states (the critical points) in the energy-wave-vector space, are proposed to fully comprehend the main features of valence and conduction bands. A detailed comparison of the sliding bilayer silicenes and graphenes is made. The theoretical predictions on the geometric structures, valence bands, and density of states could be, respectively, verified from the high-resolution measurements by scanning tunneling microscopy/tunneling electron microscopy [STM/TEM; \cite{NC7;10657,Carbon136;255, JCP129;234709, NANOLETT12;4635, NANOSCALE8;9488}], angle-resolved photoemission spectroscopy [ARPES; \cite{PRL98;206802, PRL110;146802, NP3;36, SCI313;951, PRB88;155439, SA2;1600067, SciRep7;44400}], and scanning tunneling spectroscopy [STS; \cite{APL107;263101, PCCP17;2246, SR4;7543, PRL106;126802, NP6;109, PRB95;155428, PRB77;155426, PRB87;165102, ACSNANO9;5432}.

\section{Methods}

Computer modeling and simulation serve as powerful methods for the researches of basic sciences. The numerical calculations need to exactly solve energy spectra and wave functions of quasiparticles in condensed-matter systems simultaneously. All the quasi-particles frequently experiences the distinct scattering events through the wave pocket forms, e.g., the electron-crystal-potential and electron-electron Coulomb interactions. It will be very difficult to deal with the Schrodinger equation with the single- and many-particle interactions. Some approximate models are developed to achieve the reliable solutions. Among of them, the first-principles calculations are a dominating method in the quantum mechanical simulation of periodic systems. Up to now, they are suitable and efficient in solving the very complicated intrinsic interactions under a series of approximations and simplifications to obtain the eigenvalues and eigenfunctions, and thus the reliable geometric, electronic, and magnetic properties.

For recently decades, Vienna ab initio simulation package [VASP; \cite{PRB54;11169}] computes an approximate solution to the many-body Schrodinger equation, either within the density functional theory [DFT], solving the Kohn-Sham equations \cite{JCP121;3425}, or within the Hartree-Fock approximation, evaluating the Roothaan ones \cite{PRA10;584}. The calculated results cover the optimal geometries of regular lattices/Moire superlattices, energy bands, wave-vector-dependent wave functions, spatial charge densities $\&$ their differences with independent/isolated subsystems, van Hove singularities in density of states, net magnetic moments, and ferromagnetic/anti-ferromagnetic/nonmagnteic spin configurations. Also, the other numerical methods are very powerful for the specific topics, e.g., molecular dynamics \cite{PRB47;558} and quantum Monte Carlo approaches \cite{RMP73;33}, respectively, in simulating the dynamic growth/the structural transformation and the charge-density-induced exchange and correlation interactions of emergent materials.

All the quasiparticle properties, such as energy bands, decay rates, electronic excitations, and screening charges $\&$ effective Coulomb potentials, are mainly determined by their charges and spins. Most of them could be solved by the static and time-dependent first-principles methods, especially for the former [e.g., VASP] in the geometric structures, electronic properties and spin configurations. In this work, the rich essential properties of silicene-related materials at zero temperature are thoroughly explored by the first-principles density-functional theory [DFT] under the VASP simulation \cite{PRB54;11169}. Furthermore, the local density approximation [LDA] is chosen for the model studies. The projector augmented wave approach [PAW; pseudopotentials in \cite{PRB50;17953,PRB59;1758}] is available in characterizing the electron-ion crystal potential. Very important, the exchange-correlation energy, being associated with the electron-electron Coulomb interactions, is evaluated from the Perdew-Burke-Ernzerh of functional through the generalized gradient approximation [using quantum Monte Carlo to get the charge-density-dependent many-particle energy in \cite{RMP73;33}]. Specifically, the spin-dependent single- and many-particle interactions need to be taken into account for the significant materials/adtoms, such as, the spin-orbital couplings in Si- \cite{PRL111;066804,JPCM25;395305}, Ge- \cite{JPCM25;395305}, Sn- \cite{JPCM25;395305,NJP16;105007}, Pb- \cite{PRB75;195414}, GaAs- \cite{PRL68;106}, Sb- \cite{SCR5;16108}, and Bi-induced \cite{PRB75;195414} monolayer systems. Apparently, the Hubbard-like on-site Coulomb interactions might play important roles in determining the total ground state energies, optimal geometries, band structures, Van Hove singularities, atom- $\&$ boundary-induced magnetic configurations, and magnetic moments. In monolayer silicene/graphene, a vacuum space of 15 ${\AA}$ is inserted between periodic images to avoid the interactions of neighboring cells. The cutoff energy of Block wave functions, which are built from the complete plane waves, is set to be $\sim$500 eV. The semiempirical DFT-D2 corrcetion of Grimme is calculated for the van der waals interactions. The total energy energy is defined as $E_{(DFT+D)}$=$E_{(KS-DFT)}$-$E_{disp}$, where E$_{(KS-DFT)}$ is the self-consistent Kohn-Sham energy and Edisp is a semiempirical dispersion correction. To reach the optimal geometry with the lowest ground state energy, the whole atoms of a condensed-matter system will experience the delicate relaxation processes by adjusting their positions. And then, the most stable configuration is utilized as an initial input  for the further calculations of the fundamental physical properties. For the specific calculations of electronic properties and optimal geometric structures on monolayer silicene, the first Brillouin zones are, respectively, sampled by 100$\times$100$\times$1 and 12$\times$12$\times$1 ${\bf k}$-points via the Monkhorst-Pack scheme. For orbital-project DOS calculation is given by 150x150x1. The convergence of the Helmann-Feymann force on each atom is about 0.01 eV/$\AA$.A, being accompanied with the total energy difference of ${\Delta\,E<10^{-5}}$ eV.



\section{Geometric Structures}

Monolayer group-IV systems, as clearly shown in Fig. 1, present the stable honeycomb lattices under the freestanding cases. Such materials can provide the full information about the $\pi$, sp$^2$ and sp$^3$ bondings, respectively, due to the p$_z$, [s, p$_x$, p$_y$], and [s, p$_x$, p$_y$, p$_z$] orbital hybridizations. The main features, planar or buckled structures, ground state energies, lattice constants, and heights of A and B sublattices, are consistent with the significant chemical bondings of the outmost [s, p$_x$, p$_y$, p$_z$]-orbitals. Only graphene consists of the normal hexagons [Fig. 1(a)], while the other four monoelements possess the different heights in the A and B sublattices [height differences of (0, 0.48, 0.69, 0.84, 0.89) $\AA$ for (C, Si, Ge, Sn, Pb) in Table 1]. Most important, these two sublattices are equivalent on the ${(x, y)}$-plane for any essential properties of all the group-IV materials. This result clearly illustrates that the dominance of ${sp^2}$ bonding is gradually reduced/enhanced in the increase/decrease of atomic number, while the opposite is true for the importance of ${sp^3}$ bonding. Furthermore, the sp$^2$-bonding strength is much stronger than that of the ${\pi}$ bonding. As a result, C- and Pb-related systems have the lowest/shortest and highest/longest ground state energy/lattice constant [the calculated values in Table 1], where the latter is defined as ${\sqrt 3\, b}$ and $b$ is the X-X bond length. In other words, layered graphenes are most stable among the group-IV systems, clearly indicating the best mechanical features \cite{JMCC4;5434, CSR47;6370, JPCC116;8271}. In short, the multi-orbital hybridizations fully dominate the optimal geometric structures and thus greatly enrich the fundamental properties.

The sliding bilayer graphene/silicene materials are created by the relative shift between two subsystems, where $\delta$ is in unit of lattice constant [${\sqrt 3\,b}$ $\&$ $b$=1.42/2.28 $\AA$]. The typical sliding way is chosen as follows [Fig. 2]: (a) ${\delta_a\,=0}$ [AA-bb], (b) ${\delta_a\,=1/8}$, (c) ${\delta_a\,=3/8}$, (d) ${\delta_a\,=4/8}$, (e) ${\delta_a\,=6/8}$, (f) ${\delta_a\,=8/8}$ [AB-bt], (g) ${\delta_a\,=11/8}$ $\&$ (h) ${\delta_a\,=12/8}$ along the armchair direction, and then (i) ${\delta_z\,=1/8}$ $\&$ (j) ${\delta_z\,=3/8}$ along the zigzag direction. The "b" means bottom and the "t" means top. There are four carbon/silicon atoms in a unit cell of each stacking configuration. They have the lower ground state energies /[${-20.30}$ eV]-[${-20.05}$ eV] per unit cell in Table 2, compared with that [/${-9.57}$ eV per unit cell] of monolayer graphene/silicene [Table 1]. This clearly indicates the significant attractive forces between two monolayer subsystems.

Among the distinct stacking configurations for graphene/silicene, the AB-bt/${\delta_a\,=3/8}$ one shows the lowest total energy, suggesting the most stable geometric structure. In the latter, the height differences $\Delta d$ between the intralayer A and B sublattices are greatly enhanced from 0.48 $\AA$ [Table 1] to 0.66-0.88 $\AA$. Moreover, the interlayer distance $\Delta z$ of $\sim$ 2.43-2.72 $\AA$ is shorter than  3.52 $\AA$/3.26 $\AA$ in the AA-/AB-stacked bilayer graphene.

The buckled angle, which is defined as that between the position vector of the nearest A $\&$ B sublattices e${\bf R_{AB}}$ and $x$-axis, reveals in the range of ${15.43\le\theta\le21.89}$ in Table 2. As to all the buckled honeycomb lattices, the AB-bt and AA-bb stacking configurations, respectively, presents the lowest and highest bucklings. In short, the increased buckling and the reduced interlayer distance in bilayer silicene systems are expected to induce more prominent multi-orbital hybridizations of Si-[3s, 3p$_x$, 3p$_y$, 3p$_z$].

\section{Monolayer C-, Si-, Ge-, Sn- and Pb-based materials}

Group-IV monolayer materials exhibit the diverse essential properties, mainly owing to the buckled honeycomb structures, the multi-/single-orbital hybridizations and the significant spin-orbital couplings. The height difference of A and sublattices, lattice constant, and ground state energy become large with the increase of atomic number. Roughly speaking, there are three kinds of band structures, respectively corresponding to C-, [Si, Ge]-, and [Sn. Pb]-related systems. The spatial charge density distributions and their variations are able to thoroughly examine the well-behaved $\pi$, $\sigma$ and ${sp^3}$ chemical bondings. Whether the former two are orthogonal to each other  depends on the planar or buckling geometries. All the low-lying band structures of group-IV monolayer systems could be well characterized by the tight-binding model through the fitting with the first-principles calculation results. The above-mentioned features are consistent with the spatial charge distributions and the van Hove singularities.

\subsection{The electronic properties}

Monolayer group-IV systems exhibit the diversified valence and conduction bands, as obviously illustrated in Figs. 3(a)-3(e). All the energy bands, which consist of the spin-up- and spin-down-combined states with the double degeneracy, could be classified into three kinds according to the effects of sp$^3$ bondings and spin-orbital couplings. First, graphene has a low-lying $\pi$ and $\pi^\ast$ bands in the energy range ${-3}$ eV${\le\,E^{c.,v}\le\,3}$ eV (Figs. 3(a) and 3(f)), being asymmetric about the Fermi level [${E_F=0}$] as a result of the warping effect [the non-orthonormal property of the A and B-sublattice of 2p$_z$-orbital tight-binding function; \cite{PRB78;205425}]. The linear energy dispersions just intersect at the Dirac point [the K/K$^\prime$ point] as a result of the hexagonal symmetry \cite{NAT438;201, NP7;953}. Furthermore, this Fermi-moment state only makes zero contribution to density of states [discussed later in Fig. 5]. Apparently, graphene is a zero-gap semiconductor, as directly verified from STS measurements \cite{APL107;263101, PRL106;126802, NP6;109, PRB95;155428, PRB77;155426, PRB87;165102,ACSNANO9;5432}. The low-energy essential properties are dominated by the $\pi$ bondings of 2p$_z$ orbitals [discussed later in Fig. 5]. When the valence/conduction state energies become deeper/higher, the concave-downward/concave-upward Dirac cone changes into parabolic dispersions along the direction of KM, forms a saddle structure centered at the M point, and then reaches the extreme $\Gamma$ point [not shown]. But for the direction of K$\Gamma$, it is monotonously transformed into the parabolic $\Gamma$ valley. The K and $\Gamma$ valleys are very stable in initiating the fundamental physical properties, e.g., the initial Landau levels from them, with the totally different behaviors \cite{PRB84;125455, SR5;12295}. Specifically, the K, M and $\Gamma$ points belong to the distinct critical points in the energy-wave-vector space; therefore, they will induce the different Van Hove singularities in density of states. In addition, the $\pi$-electronic state energy at the M point [${\sim\,-2.5}$ eV] is helpful in determining the strength of the nearest hopping integral for the tight-binding model \cite{NP6;109, PRB95;155428}. As to the [2p$_x$, 2p$_y$] half-occupied orbitals, they create two $\sigma$ parabolic bands which is initiated from the $\Gamma$ valley at ${E^v\sim\,-3.02}$ eV. Such deeper electronic states arise from the stronger $\sigma$ bondings, compared with the zero-energy ones of the $\pi$ bonding. The almost direct crossing between the $\sigma$ ad $\pi$ bands occur between the $\Gamma$ and K valleys along any directions, further illustrating the orthogonal feature of their chemical bondings. It should be noticed that the specific $\sigma$ band associated with the 2s orbitals come to exist at a very deep-energy range [${E^v<-6}$ eV] under the largest ionization energy [the lowest one-site energy in the tight-binding model \cite{NJP11;095003}.

There exist certain important similarities or dissimilarities between silicene/germanene [Fig. 3(b)/Fig. 3(c)] and graphene [Fig. 3(a)]. The low-lying valence and conduction bands of silicence/germanene are dominated by the $\pi$ bondings of 3p$_z$/4p$_z$ orbitals. However, the low-energy ranges of ${|E^{c,v}|}$, without the $\sigma$ bands, are, respectively, ${\sim\,3.0}$ eV, 1.0 eV and 0.5 eV for C-, Si- and Ge-related materials. This indicates the reduced $\pi$ bondings under the larger lattice constants/the wider orbitals. The Dirac-cone structure presents a slight separation in the second and third systems [insets in Fig. 3(b) and 3(c)]. A narrow band gap, ${E_g\sim\,1.55}$ meV/23.9 meV for silicene/germanene, is attributed to the enhanced spin-orbital couplings in the buckled honeycomb lattice. This is comparable to the tight-binding model with the nearest-neighbor hopping integrals and the next-nearest-neighbor spin-orbital interactions [$\lambda$; ${E_g=2\lambda\,}$ in \cite{JPCC119;11896, PCCP17;26443}] (Figs. 3(g) and 3(h)). Therefore, the single-p$_z$-orbital hybridization is roughly suitable for the low-energy physics of monolayer graphene, silicene and germanene. On the other hand, the energy spacing between the initial $\sigma$ bands and the Fermi level at the $K$ valley becomes small in the increase of atomic number, mainly owing to the enhanced sp$^3$ bonding and spin-orbital coupling. Furthermore, the direct  mixings [the anti-crossing behaviors] of $\pi$ and $\sigma$ bands might frequently occur along any directions related to the $\Gamma$ point, e.g., those along M$\Gamma$ and K$\Gamma$.

As for Sn- and Pb-related monolayer systems [Figs. 3(d) and 3(e)], the significant sp$^3$ bonding and spin-orbital coupling play critical roles in the low-lying energy bands. Apparently, two $\sigma$ bands and one $\pi$ band come to exist near the Fermi level simultaneously, in which they are, respectively, initiated from the $\Gamma$ and K valleys. The former present the splitting phenomena, being attributed to the sufficiently strong spin-orbital interaction. The semiconducting or semimetallic behaviors are co-dominated by them. Moreover, there exist the more serious band mixings along M$\Gamma$ and K$\Gamma$. The above-mentioned features clearly illustrate that the single-orbital $\pi$ bonding is not suitable for the low-energy essential properties of monolayer Sn and Pb systems, and the sp$^3$ orbital hybridizations need to be included in the phenomenological models [Figs. 3(i) and 3(j)] [ calculation details in Ref. \cite{PRB94;045410}. In short, the distinct strengths of sp$^3$/sp$^2$ bonding and spin-orbital coupling have diversified band structures and thus the other fundamental properties.

The high-resolution ARPES measurements are the only delicate ways  in efficiently examining the theoretical predictions on the occupied electronic states of 2D group-IV materials, especially for the diverse wave-vector dependences. They have verified the linear energy dispersions of the K/K$^\prime$ valleys for monolayer graphene \cite{SR4;7509}, and even under the $n$- or $p$-type dopings [the blue or red shift of the Fermi level relative to the Dirac point; Refs. []]. The diversified band structures are also identified from the stacking- and layer-number-dependent few-layer graphene systems \cite{JPCM25;085508, PRB98;195442, NANOLETT9;2654, PRB73;144427, NANOLETT12;4635, PRB78;205425}. The separated valence and conduction Dirac cones, the multi-Dirac-cone structure, the vanishing graphene-like energy bands, and the quasi-hole  semimetallic energy spectra are, respectively, revealed in silicene/Ag(111) $\&$ silicene/Ag(110)\cite{RSCA5;65255, JMCC5;627}, germanene/Au(111) \cite{PCCP19;18580, PCCP17;19039} silicene/ZrB$_2$(0001) \cite{PRB90;075422, PRB88;165404}, and tinene/Bi$_2$Te$_3$(111) $\&$ Pb/Au(111) \cite{PRB90;075105, TSF518;57, PRB87;115138}. The experimental examinations on  the $\pi$ band  at the K $\&$ M points, the $\Gamma$-valley $\sigma$ bands with the double degeneracy, their obvious mixings along K$\Gamma$ $\&$ M$\Gamma$ [Figs. 3(a)-3(e)] can provide the sufficient information about the dominance of sp$^2$ or sp$^3$ bonding, the significance of spin-orbital coupling, and their relations with the buckled/planar honeycomb lattices.

Very important, the spatial charge distributions are available in identifying/examining the multi-/single-orbital hybridizations due to various chemical bonds \cite{PCCP17;26443, PCCP19;20667, Carbon120;209, PCCP18;4000}. Both the total charge density distribution and its variation, ${\rho}$ and ${\Delta\rho}$ in Figs. 4(a)-4(j) with the ${[x, z]}$-plane projection, are delicately calculated from the first-principles calculations, in which the latter is obtained by subtracting the superposition of that in an isolated atom. Apparently, a monolayer graphene presents the $\pi$ and $\sigma$ chemical bondings [the green and red rectangles; Fig. 4(a)], respectively, arising from the ${2p_z}$ and ${[2s, 2p_x, 2p_y]}$ orbitals. Such results are further illustrated by the charge variation after the crystal formation [Fig. 4(b)]: the charge density is obviously enhanced between two carbon atoms [the red color due to the $\sigma$ bonding], while it is reduced above/below each carbon-atom [the deep blue color associated with the $\pi$ bonding]. Apparently, there exist the highest charge densities between two identical carbon atoms on the ${[x, y]}$-, ${[x, z]}$- and ${[y, z]}$-planes, being responsible for the creation of $\sigma$ bonding. Specially, graphene has the largest valence electron density within the shortest C-C bond length, compared with the other group-IV systems [Figs. 4(c)-4(j)]. The carbon honeycomb lattice has clearly displayed the best mechanical response under the lightest case \cite{RSCA4;28987}; therefore, it is very successful in the applications of material engineering \cite{RSCA4;28987}. The very strong $\sigma$ bondings are directly reflected in their collective excitation modes; that is, the $\sigma$-electronic plasmon modes have the higher oscillation frequencies, e.g., the $\sigma$ optical plasmons of graphene with frequencies higher than 10 eV in EELS and optical measurements \cite{ACSNANO5;7640, JMCC3;7632}. In addition to the [x,y]-plane three-orbital carriers, the electronic valence states, which arise from the 2p$_z$ orbitals, are extended along the $z$-direction to generate the effective distribution width. $W_z$ is estimated to be ${\sim\,0.35}$ $\AA$ from the ${[x, z]}$- and ${[y, z]}$-plane projections \cite{RSCA7;16801}. The free extension of the $\pi$ bonding on the [x, z] and [y, z] planes is the critical picture for the low-energy essential properties of graphene systems, according to the consistent theoretical and experimental studies, e.g., the gapless and semiconducting Dirac-cone band structure \cite{PSS244;4106}, the doping-induced Landau dampings and 2D plasmons \cite{NJP12;033017, JPCM23;012001}, the unusual magneto-electronic Landau levels \cite{PCCP17;26008}, the specific magneto-optical selection rules \cite{PCCP17;26008, PCCP18;17597, ACSNANO4;1465, SR8;13303}, the rich magnetoplasmon modes \cite{NANOSCALE6;10927, PRB99;195447}, and the unique quantum Hall conductivities \cite{PRB99;085443, PRB97;085413}.

Both $\pi$ and $\sigma$ chemical bondings, which survive in the  Si-/Ge-/Sn-/Pb-related monolayer materials, become non-orthogonal to each other, as clearly observed in Figs. 4(c)-4(j). Their bonding strengths decline as the atomic number of group-IV materials grows. This leads to the longer bond lengths [Table 1]. However, the opposite is true for the observable/significant sp$^3$ bonding in the buckled systems. The effective distribution widths related to the 3p$_z$, 4p$_z$, 5p$_z$, and 6p$_z$ orbitals are, respectively, estimated to be about 0.37, 0.39, 0.41 and 0.45 $\AA$'s, as obtained from the ${[x, z]}$- and ${[y, z]}$-plane projections. They are roughly consistent with the probability distributions of the isolated atomic orbitals. The total charge density can provide the information of the A- and B-sublattice height difference [Table 1], being enhanced in the increase of atomic number. The large buckling indicates that it is relatively difficult to synthesize few-layer group-IV materials with heavy masses. To fully comprehend the essential properties, the non-negligible ${sp^3}$ multi-orbital hybridizations, as well as the spin-orbital coupling, might have to be included in the tight-binding model simultaneously \cite{JPCC119;11896, SR9;36547, SR7;12069, PCCP20;11369}.

The four-orbital-decomposed density of states is very useful in understanding the strong effects of multi-orbital hybridizations on essential properties. The various band-edge states of electronic energy spectra, as clearly illustrated by density of states in Figs. 5(a)-5(e), can create the diverse 2D van Hove singularities. Generally speaking, the critical points in energy-wave-vector space, which appear in energy bands of Figs. 3(a)-3(e), cover the Dirac-cone bottoms, the extreme states of parabolic dispersions, the saddle points, and the constant-energy loops. Such electronic states, respectively, generate the V-shape structure, the shoulders, the logarithmically symmetric peaks, and the square-root-form asymmetric ones. For a pristine monolayer graphene [Fig. 5(a)], its unusual density of states presents a linear energy dependence being symmetric about the Fermi level, the $\sim$${-2.5}$ eV $\pi$ peak/the ${\sim\,1.90}$ eV $\pi^\ast$ peak, and the ${\sim\,-3.02}$ eV shoulder arising from the ${2p_x}$ $\&$ ${2p_y}$ orbitals. The energies of the symmetric $\pi$ peak and the $\sigma$ shoulder increase as the atomic number grows [Figs. 5(a)-5(e)]. The critical mechanisms are the reduced $\pi$ bonding and the enhanced dominance of sp$^3$ bonding and spin-orbital coupling. The strong coupling of chemical and physical interactions further lead to the dramatic transformation of one${\rightarrow}$two shoulder in Sn and Pb [Figs. 5(d) and 5(e)]. There exists an asymmetric peak after the formation of $\sigma$ bands, clearly indicating the mixing of  orbital-dominated energy bands, such as, the ${\sim\,-1.52}$ eV, ${-1.4}$ eV, ${-1,3}$ eV, and ${-1.2}$ eV peak structures for [Si, Ge, Sn, Pb], respectively. This indicates the existence of the significant ${sp^3}$ bonding. The examinations of the STS measurements on monolayer group-IV materials are able to identify the main features for the $\pi$, sp$^2$ $\&$ sp$^3$ bondings, as well as the spin-orbital interaction.

\subsection{Significant differences among monolayer group-IV systems}

Monolayer group-IV materials, as clearly indicated in the previous section, exhibit the diverse low-lying energy bands because of the buckled/planar honeycomb lattices, the multi-/single-orbital hybridizations, and the significant/negligible spin-orbital couplings. The generalized tight-binding model \cite{SR8;11070, SR9;36547, SIAM56;499, JPCC119;11896, SR7;12069, PRB77;045407, PRB82;245412, PCCP20;11369}, with the reliable parameters in the presence of electric and magnetic fields, could be further developed to fully explore the diversified essential properties through the direct combinations with the single- \cite{PRB74;085406} and many-particle theories \cite{PRB62;8508}, such as, optical absorption spectra $\&$ selection rules \cite{PCCP17;26008, PCCP18;17597}, magneto-electronic properties \cite{ACSNANO4;1465, SR8;13303}, Coulomb excitations and decay rates \cite{SR7;40600, PRB98;195442, PRB74;085406}, and quantum Hall effects \cite{NAT438;201, NP7;953, PRB99;085443, PRB97;085413}. For example, a buckled tinene honeycomb lattice, which possesses the strong sp$^3$ chemical bonding and thus creates the multi-degenerate energy bands [Fig. 3(d)].  The magnetically quantized energy spectra and the sublattice-, orbital- and spin-dependent wave functions [subenvelope functions] could be computed very efficiently through the exact diagonalization method even for a rather large complex Hamiltonian matrix. Furthermore, the chemical and physical pictures are responsible for the diversified magneto-electronic properties [also for the magneto-optical absorption spectra and selection rules; \cite{PCCP17;26008, PCCP18;17597, ACSNANO4;1465, SR8;13303}]. Recently, the calculated results show that tinene presents the unusual magnetic quantization \cite{SR7;1849}, being absent in graphene, silicene and germanene. The multi-valley electronic structure, which correspond to the low-energy $\pi$ and $\sigma$ energy bands, is capable of generating two groups of low-lying Landau levels [LLs], with the different orbital components, spin configurations, localization centers, state degeneracy, and magnetic- and electric-field dependencies. The first and second groups are dominated by the 5p$_z$ and [5p$_x$,5p$_y$] orbitals, respectively. The LL splittings in the first and second groups are due to the coupling effects of the perpendicular electric and magnetic fields, respectively. Specifically, the LL anti-crossings only appear in the first group during a variation of the electric field. The inter-atomic spin-orbital coupling between 5p$_z$ and [5p$_x$,5p$_y$] orbitals and the Coulomb potential difference between A and B sublattices result in the dramatic probability redistribution of the opposite spin components and thus the frequent LL anti-crossing behaviors. The similar magneto-electronic diverse phenomena are expected to be revealed in monolayer Pb. On the other hand, monolayer graphene, silicene and germanene present the totally different phenomena, mainly owing to the $\pi$-bonding-dominated low-energy physics. There are only one groups of well-behaved valence and conduction LLs, in which each localized oscillation mode has a specific zero-point number. When the buckled silicene and gemanene are in the presence of gate voltage, electronic states/magneto-electronic ones are split by the cooperation of the spin-orbital coupling and the sublattice-dependent Coulomb potentials, as observed in the tinene and Pb systems. There exist the valley-split or spin-split LL energy spectra, being very sensitive to the strength of electric and magnetic fields. The theoretical predictions on the diverse magneto-electronic properties in 2D group-IV materials could be verified from the high-resolution measurements of magnetic van Hove singularities \cite{NP6;109, PRB95;155428}, quantum Hall transports \cite{NP7;948}, and magneto-optical absorption spectra \cite{PCCP17;26008, PCCP18;17597, ACSNANO4;1465, SR8;13303}. In short, [Sn, Pb]-, [Si, Ge]- and C-related monolayer materials have the different electronic properties, and so do the other fundamental properties.

\section{Stacking-configuration-enriched essential properties in bilayer graphenes and silicenes}

For bilayer graphene systems, their interlayer attractive forces are clearly identified to be due to the van der Waals interactions of pure carbon atoms \cite{DRM88;151, PRB85;245430}. The $\sigma$ bondings, which are generated by [2s, 2p$_x$, 2p$_y$] orbitals, only lie on the [x, y]-plane honeycomb lattice. Their valence energy spectra, with ${E^v<-3.0}$ eV, are hardly affected by bilayer couplings, and so do the higher-/deeper-energy physical properties. The weak, but significant interlayer atomic interactions mainly come from the $\pi$ bondings of C-2p$_z$ orbitals. The interlayer hopping integrals are much smaller than the intralayer one, while the low-energy physical properties are closely related to their stacking dependencies. That is to say, the single 2p$_z$-orbital Hamiltonian \cite{PRB82;245412, PRB85;115423} is sufficient/reliable for the stacking-enriched phenomena in few-layer graphene systems. The phenomenological models are very successful in exploring the other essential properties, such as, the rich optical properties \cite{SR8;11070, NJP15;015010, RSCA4;63779}, Coulomb excitations \cite{SR7;40600, PRB98;195442}, and inelastic scattering rates \cite{PRB83;155441}. On the other side, the sliding bilayer silicenes have the larger height differences between the intralayer A and B sublattices/the smaller interlayer distances, compared with monolayer silicene/bilayer graphenes. These two critical factors will enhance sp$^3$ bondings and greatly enlarge the interlayer hopping integrals, in which the latter even dominate the kinetic energies of the intrinsic Hamiltonian \cite{SIAM56;499}. Consequently, the $\pi$ and $\sigma$ bands are difficult to distinguish from the complicated electronic energy spectra. Up to now, few calculations of the tight-binding model, being focused under the specific AA-bt and AB-bt bilayer silicenes \cite{PRB97;125416}, are only conducted on the 3p$_z$-orbital interactions. As a result, they are useful in understanding the unusual magnetic quantization phenomena and optical absorption spectra due to the first pair of valence and conduction bands. Even for well-defined stacking configurations in bilayer silicenes, it is very difficult to thoroughly simulate the low-lying electronic structures with the strong wave-vector dependencies in the first-principles calculations. This has created high barriers in the near-future studies of phenomenological models for few-layer silicene/germanene/tinene systems \cite{SR6;20714, NP14;344, JPCM25;085508, NANOSCALE8;9488, CM30;4831, CPB24;086102, PCCP18;18486, JAP114;094308}.

\subsection{Sliding bilayer graphene}

The optimal geometric structures and electronic properties of the configuration-dependent bilayer graphene systems are thoroughly explored by using the first-principles method \cite{PRB84;113412, RSCA4;63779, JPCC119;10623, EPJ148;91}; also sees the tight-binding model in \cite{SR8;11070, SIAM56;499, PRB77;045407, PRB82;245412}. The geometric structures for sliding bilayer graphene systems are illustrated under the specific path in Fig. 2. All carbon atoms possess the identical environment for AA$^\prime$ stacking (Fig. 2(f)), while their (x, y)-projections are different on the first and second layers. Any stacking configurations present the planar honeycomb lattices, clearly illustrating that the dominant $\sigma$ bondings on the (x, y) plane are hardly affected by them. Obviously, the interlayer distance, as indicated in Table 2 by $\triangle d$, declines and then grows during the variation of the relative shift between two layers. Furthermore, it is shortest for the AB stacking (3.26 $\AA$), but longest under the AA stacking (3.53 $\AA$). This result is consistent with the TEM measurements (3.35{\AA} and 3.55{\AA}, relatively, for AB and AA stackings; \cite{JCP129;234709}) and other theoretical predictions \cite{NanoT21;065711, RSCA3;3406}. The total ground state energy is strongly affected by the relative shift (Table 2). As a result, the shortest interlayer distance creates the strongest van der Waals interactions among the 2p$_z$ orbitals. The AB stacking is thus expected to be the most stable configuration of bilayer graphene systems, being frequently revealed in experimental samples \cite{JCP129;234709, PRB48;17427}. The stacking-induced difference of ground state energy is about 30 meV, in which it is one order smaller than that in bilayer silicene systems (Table 2). Such result clearly illustrates more strong interlayer chemical bondings in the latter.

Apparently, the calculated results, as illustrated in Figs. 6-8, demonstrate the dramatic transformations of electronic properties, especially for the sensitive dependencies of low-lying energy bands on the magnitude and direction of the relative shift between two graphene sublayers.  The $\pi$-electronic energy bands, which are generated through the AA{$\longrightarrow$}AB{$\longrightarrow$}AA$^\prime${$\longrightarrow$}AA path (Fig. 6), present the stacking-diversified phenomena. Specifically, two pairs of vertical Dirac-cone structures (two Dirac points at the K and K$^\prime$ valleys in Fig. 6(a) of ${\delta_a\,=0}$), are dramatically  transformed into parabolic bands (Fig. 6(d) at ${\delta_a\,=8/8}$), non-vertical Dirac cones (Fig. 6(f) at ${\delta_a\,=12/8}$) and the non-well-behaved electronic energy spectra. The last ones present the creation of an arc-shaped stateless region near the Fermi level (Figs. 6(b), 6(c), 6(e), 6(g) and 6(h)), respectively, at ${\delta_a\,=1/8}$, 4/8, 11/8 (armchair direction), ${\delta_z\,=1/8}$ and 3/8 (zigzag one), strongly distorted energy dispersions, significant overlaps of valence and conduction bands, extra low-energy saddle points, and the splitting of  saddle M-point states (also in AA$^\prime$ stacking of Fig. 6(f)). The $\pi$-band widths, which are associated with the KM$\Gamma$/KM$^\prime$$\Gamma$ composite structures, can reach the range of ${\sim\,7-8}$ eV for any stacking configurations. They have been extended by the interlayer hopping integrals, compared to that of monolayer graphene \cite{NRL9;110}. Concerning the $\sigma$ bands due to (2p$_x$, 2p$_y$) orbitals, they are initiated from the stable $\Gamma$ valleys, and their state degeneracy becomes double. Apparently, the interlayer atomic interactions hardly depend on the C-(2s, 2p$_x$, 2p$_y$) orbitals in two graphene subsystems. Such degenerate energy dispersions also present the saddle M-/M$^\prime$-point structure and terminate at the stable K/K$^\prime$ valleys; therefore, their widths are more than 4 eV. Among all the bilayer graphene systems, the AA stacking has the highest free carrier density, in which band overlap even reaches ${\sim\,1}$ eV (the largest density of states at the Fermi level; \cite{JPCC119;10623}). In short, the feature-rich band structures of bilayer graphene systems, which, cover the highly anisotropic energy dispersions, the dramatic transformations/the drastic changes of Dirac-cone structures near the stable K/K$^\prime$ valleys, the extra low-lying saddle points, the eye-shape stateless region,  the split saddle M $\&$ M$\prime$ points in non-AA $\&$ non-AB stackings, the degeneracy-enhanced $\sigma$ bands, and the direct crossings of $\pi$ $\&$ $\sigma$ bands, could be directly examined by the high-resolution ARPES measurements. Furthermore, the M and M$^\prime$ saddle points are expected to cause very strong absorption peaks at about 4-5 eV due to the high concentration the $\pi$-electronic states.

The spatial charge densities in sliding bilayer graphene systems present the observable and significant variations, mainly owing to the interlayer van der Waals interactions. The various ${\Delta\rho}$'s are clearly revealed in Figs. 7(a)-7(h), depending on the various stacking configurations. Apparently, the $\pi$ and $\sigma$ chemical bondings remain the well-defined forms; that is, they are easily identified from the carrier distributions between two neighboring carbon atoms. They are orthogonal to each other on the planar honeycomb lattices. The latter almost keep the same after the bilayer couplings; therefore, the $\sigma$ valence bands only exhibit the rigid shift (Figs. 6(a)-6(h)). Only the former are somewhat modified under the attractive interlayer atomic interactions. The effective distribution width of ${2p_z}$ orbitals becomes wider, compared with that in monolayer systems. The interlayer hopping integrals related to them are sensitive to the stacking symmetries and responsible for the diversified physical phenomena, such as, the dramatic transformation of band structures/van Hove singularities during the variation of bilayer stacking configurations (Figs. 6(a)-6(h)/8(a)-8(h)). The clear separation of $\pi$ and $\sigma$ chemical bindings, as well as the former's dominance at the low-energy properties, could be found in all the ${sp^2}$-based carbon materials, e.g., the AA- \cite{JES148;1159}, AB- \cite{EPJ148;91} $\&$ ABC-stacked \cite{ACSNANO6;5680}, bulk graphites \cite{JES148;1159}, layered graphene systems with AA, AB \cite{PRB84;113412, EPJ148;91, JES148;1159}, ABC \cite{PRB94;165428, PRB84;125455}  $\&$ AAB stackings \cite{PRB98;195442}, achiral $\&$ chiral carbon nanotubes \cite{JPSJ71;1820, PRB50;17744, PRB61;2468}, and graphene nanoribbons \cite{NRL10;254, JMCA1;10762, JAP113;183715}. The up-to-date tight-binding model, which is built from the ${2p_z}$-orbital periodical functions \cite{JPCC119;11896, SIAM56;499}, is capable of fitting the first-principles low-lying energy bands. And then, it is very useful in creating and understanding the diversified essential properties in graphene-related emergent materials \cite{PCCP19;20667, Carbon120;209, PCCP18;4000, PRB92;075403, SSS50;69}, such as, the rich and unique magnetic quantization phenomena with the strong dependences on stacking configurations \cite{NM12;887892, NANOLETT9;2654, SR4;7509, SCI313;951954}, layer numbers \cite{PRL98;206802, PRB88;155439}, dimensions \cite{PRL98;206802, PRL110;146802, NP3;36, ACSNANO6;6930, PRB73;045124, PRL100;037601}, field modulations \cite{Carbon99;432}, composite fields \cite{PRB94;045410, APL105;222411}, free carrier dopings \cite{PRB73;144427, JPSJ76;024701, PRL98;166802, PRB77;085426, RMP83;1193, PRL99;216802}.

Densities of states in sliding bilayer graphenes exhibit a lot of unusual structures due to the rich band-edge states, as clearly shown in Figs. 8(a)-8(h). The van Hove singularity at the Fermi level presents the plateau (Figs. 8(a), 8(e), 8(f) $\&$ 8(g)) or dip structure (Figs. 8(b), 8(c), 8(d) $\&$ 8(h)), respectively, associated with the Dirac-cone-like (Figs. 6(a), 6(e), 6(f) $\&$ 6(g)) and parabolic bands (Figs. 6(b), 6(c), 6(d) $\&$ 6(h)). Both structures possess the finite values at $E_F$, clearly illustrating the semi-metallic behaviors in all bilayer graphene systems. The free carrier density shows a non-monotonous dependence on the relative shift. The critical mechanisms, the stacking-enriched interlayer hopping of C-2p$_z$ orbitals, create the significant overlaps of low-lying valence and conduction bands. The neighboring special structures of the plateaus and the dips, respectively, belong to cusps (Figs. 8(a), 8(e), 8(f) $\&$ 8(g)) and shoulders/symmetric peaks (Fig. 8(d)/8(b), 8(c) $\&$ 8(h)). The low-lying logarithmic peaks come from the extra saddle points of greatly modified energy dispersions, and their number is very sensitive to stacking configuration. Moreover, the middle-energy prominent peaks from the M and M$^\prime$ points are revealed as the single- and twin-peak structures. The intensity and energy of the prominent peaks strongly depend on their splitting and merging as the relative shift occurs. The induced saddle points, the splitting of the middle-energy peaks and free carriers at $E_F$ could be verified from the STS experiments. The low- and middle-energy saddle points and the derived prominent peaks are closely related to stacking configurations; therefore, the experimental measurements on them is available in resolving the subangstrom misalignment stackings of bilayer graphene systems \cite{RSCA4;63779, SR4;7509}.

According to the first-principles calculations \cite{DRM88;151, PRB94;165428, PRB95;075422}, bilayer graphene systems possess the low-energy $\pi$-electronic states under any stacking configurations. Their band structures could also be understood from the tight-binding model with the well-fitted intralayer and interlayer hopping integrals from the empirical interaction formula \cite{PRB82;245412}. In general, the carbon-2p$_z$ orbitals in all the sp$^2$-dominated systems are sufficient for characterizing the rich and unique phenomena, e.g., the $\pi$ and $\pi^\ast$ bands in  the AA- \cite{JES148;1159}, AB- $\&$ ABC-stacked graphites \cite{JPCM25;085508, EPJ148;91, ACSNANO6;5680, PRB82;035409}, few-layer graphene systems with AAA, ABA, ABC $\&$ AAB stackings \cite{PRB98;195442, Carbon136;255, NANOLETT12;4635, PRB77;155426, PRB87;165102, PCCP18;4000}, achiral and chiral \cite{NANOLETT5;197} carbon nanotubes, 1D graphene nanoribbons under the open/passivated edges \cite{PRB85;245430, PCCP15;868}, carbon onions \cite{NAT347;27, CPL336;201}, and ${C_{60}}$-related fullerenes \cite{PRL68;631, SCI280;1253}. This fact is confirmed by both theoretical predictions [the numerical and phenomenological methods] and experimental examinations on band structures \cite{NM9;315, NANOLETT11;4574}, magneto-electronic states \cite{SR4;7509}, absorption spectra \cite{SR3;3143, JPCL7;2328}, quantum transports \cite{NP7;948}, and Coulomb excitations \cite{PRB74;085406, PRB62;8508} $\&$ decay rates \cite{PRB90;045150}. In short, the pure few-layer graphene systems could be well understood from the calculations of phenomenological models, while the chemical modifications due to absorptions, substitutions and defects might create the structure- and bonding-dependent non-uniform environments and thus the near-future open issues \cite{Carbon136;255}.

\subsection{Sliding bilayer silicene}

The 2D band structures are remarkably diversified by the relative shift between two silicene sublayers. The electronic structure of AA-bb ($\delta_a$=0), as clearly shown in Fig. 9(b) presents six/four valence/conduction bands, being double values relative to monolayer silicene displayed in Fig. 9(a). The small overlaps of valence and conduction bands at the $\Gamma$ and M points demonstrate that silicene possess the semi-metallic behavior. The valence and conduction states hold the free holes and electrons respectively above/below the Fermi level. Besides, three valence bands and one conduction band of bilayer silicene, with the oscillatory energy dispersions, are very close to $E_{F}$. These physical phenomenon illustrates clearly the significant contributions of $\pi$ and $\sigma$ electrons for the low-energy essential properties. Specifically, the first conduction band with the partially flat energy dispersion near the M point might contribute to high density of states (discussed later in Fig. 11(b)). The constant-energy loops, being associated with the band mixings, come to exist along $\Gamma$M/$\Gamma$$M^{'}$ and KM/K$M^{'}$. The saddle M-point structures are revealed in all energy bands, thus leading to the prominent van Hove singularities. The valleys at $\Gamma$ and K points belong to the parabolic structures. Obviously, it might be meaningless in characterizing band widths of the $\pi$ and $\sigma$ bondings. The composite valley structures, similar to those of monolayer silicene (disperse along the $K\rightarrow M\rightarrow\Gamma$ and $\Gamma\rightarrow M\rightarrow K$ ones, respectively, due to the $\pi$ and $\sigma$ bands in Fig. 9(a)), cannot survive in this bilayer system. The outer four orbitals of silicon atoms need to be taken into account for the multi-orbital hybridizations simultaneously. The above-mentioned electronic energy spectrum is doubly degenerate in the absence of spin configurations. The spin-dependent interactions are fully negligible in the bilayer AA-bb stacking, mainly owing to the high-symmetry stacking and the strong inter bondings (discussed later in Fig. 10).

The electronic energy spectra are very sensitive to the shift. For $\delta_a$=1/8 in Fig. 9(c), this bilayer stacking morphology remains the semi-metallic property, while the main features of electronic structures are changed. The state degeneracy of the high-symmetry M and $M^{'}$ points are thoroughly destroyed. That is to say, an obvious change from the six-fold into two-fold rotational symmetry results in their energy splitting with a wide range of $\sim$0.5 eV. Only the saddle M-/M$^{'}$-point structures are capable of accumulating a lot of electronic states (density of states in Fig. 11(c)); they are expected to exhibit very prominent optical absorption peaks. Similar phenomenon could be also found in carbon-related sp$^2$-bonding systems, such as, layered graphenes \cite{PCCP17;26008, PCCP18;17597, NJP15;015010, APE9;065103, RSCA4;63779, APL98;261920}, carbon nanotubes \cite{JPSJ71;1820, PRB50;17744, PRB61;2468}, and graphite \cite{JAP85;7404, NJP12;083060, PR138;197, Carbon3;401, NRL10;234, Carbon5;403}.


It should be noticed that electronic states in some energy bands create the parabolic valleys near the M/M$^{'}$ points. All the band-edge states appear at the $\Gamma$, M, M$^{'}$ and K points and in between them. The similar semi-metallic behavior is revealed as the weak band overlap under $\delta_a$=8/8 without spin configuration, $\delta_a$=4/8, $\delta_a$=11/8, $\delta_a$=12/8, and $\delta_z$=3/8 (Figs. 9(h), 9(e), 9(i), 9(j), and 9(l), respectively). The other stacking configurations belong to the zero-gap (Figs. 9(f) and 9(k)) and indirect-gap semiconductors (Figs. 9(d) and 9(g)), respectively, corresponding to $\delta_a=6/8$ , $\delta_z$=1/8, $\delta_a=3/8$, $\delta_a=8/8$ with spin arrangement.

Specifically, the AB-bt bilayer silicene, as clearly shown in Figs. 9(g) and 9(h), exhibits the unusual energy bands. This system has the six-fold rotational symmetry even under the enhanced buckling, creating the degenerate M and $M^{'}$ point. In the energy range of $\mid E_{c,v}\mid$$<$ 0.5 eV, band structures, with and without spin-dependent interactions, are quite different from each other. This result clearly suggests that the spin interactions are greatly enhanced by the AB-bt stacking configuration. Its strength is roughly estimated to be $\sim$ 0.5 eV. There exists an indirect band gap of Eg $\sim$508 meV, which corresponds to the lowest unoccupied conduction state at the K point and the highest occupied valence state along the K$\Gamma$ direction. Apparently, the first pair of energy bands presents the non-monotonous dispersion relations. On the other hand, without spin interactions, there exists a weak overlap of the first pair of energy bands near the K point (along the K$\Gamma$ direction) [Fig. 9(h)]. These two bands are deduced to be dominated by the $\pi$-electronic states; that is, they are almost independent of $\sigma$ orbitals. Their low-energy physical properties hardly depend on the $sp^{3}$ bondings, e.g., the magnetic quantization \cite{PRB97;125416} and magneto-optical selection rules \cite{SR9;36547}. However, the opposite is true for the other stacking configurations.

The spatial charge density distributions present the dramatic transformations after including the interlayer atomic interactions. The bilayer silicene systems are totally different from monolayer one, as clearly observed from Figs. 10(a)-10(j). The charge density, which is distributed between two neighboring silicon atoms on the same layer (the red regions), is greatly reduced/transferred, compared to a monolayer silicene (Fig. 4(c)). The lower/upper part of the $\pi$ bonding in the first/second layer obviously show the significant transformation; that is, it is thoroughly absent. Also, more charge densities come to exist in the interlayer spacing (the light red or yellow regions). These clearly illustrate that electron carriers are transferred from the first and second layers to the interlayer spacing and there exist the important four-orbital hybridizations, i.e., (3s, 3$p_{x}$, 3$p_{y}$, 3$p_{z}$)-(3s, 3$p_{x}$, 3$p_{y}$, 3$p_{z}$) bondings. The interlayer bonding reaches the maximum in $\delta_a$=3/8 (the red region between layers). After that, the interlayer bonding strength turns to weak gradually through $\delta_a$=4/8 to $\delta_a$=12/8, and then gradually disappears along the shift of zigzag direction. The strong dependence of orbital-created intralayer/interlayer hoppings on the stacking structure is responsible for the diverse electronic properties.

The 2D van Hove singularities, which are clearly shown in Fig. 11(a)-11(l), directly reflects the primary features of electronic energy spectra (Figs. 9(a)-9(l)). Furthermore, the orbital-decomposed densities of states are very useful in fully exploring the significant contributions due to (3s, 3$p_x$, 3$p_y$, 3$p_z$) of silicene-based materials and thus distinguishing the important differences among the $\pi$, $\sigma$ and $sp^{3}$ bondings. As for a pristine monolayer silicene, its density of states is vanishing at the Fermi level $E_F$ = 0 shown in Fig. 11(a), obviously illustrating the semiconducting behavior. The 3$p_z$-orbital contributions (the blue curve) present a linear E-dependence at low energy, and a symmetric peak in the logarithmic divergent form at $E^{v}$=-1.03 eV/$E^{c}$=0.63 eV (a $\pi$/$\pi^\ast$ peak). In the region of ${E<-3.20}$ eV, the 3$p_z$ contribution vanishes. These features, which, respectively, arise from a very narrow gap, the almost isotropic linear bands initiated from the K valley, the saddle M-point structures of $\pi$/$\pi^{*}$ valence/conduction bands, and the parabolic $\Gamma$ valley, are mainly determined by the $\pi$ bondings (the valence/conduction band-edge states in the $K\rightarrow M\rightarrow\Gamma$ composite valleys in Fig. 9(a)).

The $\pi$ and $\pi^{*}$ peaks are partially contributed by the 3s, 3$p_x$ and 3$p_y$ orbitals (the red, green and brown curves). The (3$p_x$, 3$p_y$)-dominated valence van Hove singularities include an initial shoulder structure at E=-1.2 eV, a prominent symmetric peak in the logarithmic form at E=-2.70 eV, and a terminated shoulder one at E=-4.50 eV. They correspond to the first valence band along the $\Gamma\rightarrow M\rightarrow K$ direction. Apparently, the $\pi$ and $\sigma$-orbital (3$p_x$, 3$p_y$) hybridizations are characterized by two very strong asymmetric peaks at E=-2.21 eV and E=-2.72 eV. Also, the $sp^{3}$ orbital-hybridization is clearly seen at -3.2 eV. Moreover, most of 3s-orbital contributions locate in the regions of E$<$-3.0 eV and E$>$1.0 eV. The above-mentioned results have shown the separation/significant hybridizations of the $\pi$ and $\sigma$ bands in the different energy ranges.

The main features in densities of states, as illustrated in Figs. 11(b)-(l), become very complicated in the presence of interlayer couplings. Apparently, the semi-metallic and semiconducting bilayer silicene systems, which are characterized by their values near the Fermi level, are, respectively, revealed in Figs. 11(b), 11(c), 11(e), 11(h), 11(i), 11(j), and 11(l), and Figs. 11(d), 11(f), 11(g), and 11(k). The finite densities of states in the former are frequently accompanied with the strong special structures, mainly owing to the weak energy dispersions across $E_F$=0. However, the vanishing density-of-state ranges of the latter might be close to zero or observable, where the asymmetric V-shapes in Figs. 11(f) and 11(k) across $E_{F}$ corresponding to the modified Dirac-cone structures (Figs. 9(f) and 9(k)). The AB-bt system ($\delta_a$=8/8) could possess a band gap of $\sim$0.31 eV [Fig. 11(g)] or Dirac points/band overlap at the Fermi level [Fig. 11(h)] depending on whether the spin interactions are included. In the low-energy region of $|E|<0.5$ eV, only few stacking configurations are dominated by the 3$p_z$ orbitals such as, $\delta_a$=6/8 and 8/8 in Figs. 11(f) and 11(g), respectively. The $sp^3$ hybridization is observable in Figs. 11(d)-(k) according to the prominent peak at $E\approx$-5 eV. However, in general, it is very difficult to distinguish the initial, middle (saddle-point structure) and final $\pi$/$\sigma$ bands; therefore, both bilayer and monolayer silicenes [Figs. 11(b)-(l) and Fig. 11(a)] are totally different from each other. There are a lot of symmetric peaks, antisymmetric ones and shoulders in the whole energy range, mainly owing to the multi-orbital hybridizations, the oscillatory energy dispersions, and the frequent anticrossings.

\subsection{Significant differences between bilayer silicenes and graphenes}

The sliding bilayer silicene and graphene systems exhibit the diverse geometries, electronic and magnetic properties, as a result of the significant chemical bondings of outer four orbitals. The former and the latter, respectively, possess the buckled and planar honeycomb lattices, the interlayer distances of $\sim$2.43-2.72 $\AA$ $\&$ $\sim$3,26-3,53 $\AA$,  the total energy differences among the different configurations about ${0.25}$ eV $\&$ 0.03 eV, and the high and low charge densities between the interlayer spacing. Furthermore, the enhanced buckling angle/height difference between A and B intralayer sublattices can reach ${\sim\,21.89^\circ}$ and ${\sim\,0.88}$ $\AA$. These results clearly illustrate that Si-(3s, 3p$_x$, 3p$_y$, 3p$_z$) and C-(2p$_z$) play important roles in the interlayer atomic interactions, respectively, leading to the totally modified electronic properties and the drastic changes of the $\pi$-/$\pi^\ast$-band ones. The electronic energy spectra of bilayer silicenes present the thorough absence of  Dirac-cone structures, the semimetals/direct-or indirect-gap semiconductors, the multi-constant loops, the parabolic ($\Gamma$, K, M) valleys, the M- $\&$ M$^\prime$-related split energy bands except for the high-symmetry AA $\&$ AB stackings, negligible/significant spin interactions under the non-AB/AB stacking, the frequent band mixings, a lot of band-edge states, and the undefined $\pi$ and $\sigma$ band widths. On the other side, bilayer graphenes show the highly ${\bf k}$-dependent valence and conduction bands, the dramatic transformations between two pairs of vertical/non-vertical Dirac cones and parabolic bands initiated from the stable K/K$^\prime$ valleys, the sliding-induced low-energy saddle points, the eye-shape stateless region,  the split saddle M $\&$ M$\prime$ points  in non-AA $\&$ non-AB stackings, the degeneracy-enhanced $\sigma$ bands,  the direct crossings of $\pi$ $\&$ $\sigma$ bands, and the well-defined band widths. The high-resolution ARPES examinations are able to finalize the cooperative/competitive relations  between the $\pi$ and $\sigma$ bondings, and the presence/absence of the spin interactions. As to the rich and unique van Hove singularities, monolayer and sliding bilayer silicenes, respectively, present (a) the V-shape structure at $E_F=0$, the well-behaved $\pi$ $\&$ $\sigma$ bands by the initial cusp/shoulder, logarithmic peak and final shoulder, as well as two square-root asymmetric peaks during the band mixings, and (b) the finite or vanishing density of states at $E_F$/in a band-gap region, a lot of stacking-induced shoulders, symmetric $\&$ asymmetric peaks, and the undefined $\pi$ $\&$ $\sigma$-electronic structures. The STS measurements are available in distinguishing the dominance of the $\pi$ or sp$^3$ bonding on the low-energy physical properties. The above-mentioned main features of electronic properties could be well described by the four- and single-orbital hybridizations for bilayer silicenes and graphenes, respectively. As a result, it is almost impossible to simulate the former using the phenomenological models, while the opposite is true for the latter. For example, the magneto-electronic Landau levels \cite{SR4;7509}, magneto-optical absorption spectra \cite{ACSNANO4;1465, APL105;222411}, magnetoplasmons \cite{SR8;13303, PCCP15;868} and Hall conductivities \cite{PCCP19;29525}, which are revealed in sliding, twisted and modulated bilayer graphene systems, are fully investigated by the generalized tight-binding model \cite{PRB77;045407}, as well as the single- and many-particle theories \cite{PRB97;125416,ACSNANO4;1465,PCCP19;29525}.

\section{Concluding Remarks}

The theoretical framework, which is developed under the first-principles calculations, is very useful in fully understanding the diverse phenomena of essential properties in layered materials. The various 2D emergent condensed-matter systems cover the single-layer group-IV monoelements, sliding bilayer graphenes and silicenes, being accompanied with the different critical factors [buckled/planar geometries, stacking symmetries]. The delicate evaluations and thorough analyses are mainly focused on the optimal lattice symmetries,  the atom- $\&$ spin-dominated electronic energy spectra, the semiconducting/semi-metallic/metallic behaviors [the 2D carrier densities], the well-defined or undefined $\pi$- $\&$ $\sigma$-electronic energybands, the spatial charge densities $\&$ their variations after the crystal formations, interlayer couplings, the atom-, orbital- $\&$ spin-decomposed van Hove singularities, and the vanishing or finite magnetic moments. The key pictures mainly come from the single- $\&$ multi-orbital hybridizations and spin configurations. Part of calculated results are consistent with the up-to-date experimental measurements, while most of them require the further detailed examinations.

For group-IV 2D materials, STM/TEM/LEED, STS and ARPES, have been utilized to examine the optimal geometries, the van Hove singularities, and the occupied valence bands, respectively.  The top and side views of monolayer/few-layer graphene [the planar honeycomb lattices and the stacking configurations] are, respectively, verified by STM \cite{PRB79;125411} and TEM \cite{JESR195;145}. The STS examinations on van Hove singularities are only done for monolayer graphene \cite{PRL109;196802}, in which they are capable of verifying the V-shape structure with a almost zero density of state at the Fermi level, and the middle-energy $\pi$ and $\pi^{\ast}$ peaks. Such measurements on other group-IV systems are expected to be available in identifying the critical roles of sp$^3$/$\&$ sp$^2$ bondings and spin-orbital couplings. The low-energy ARPES experiments, being conducted on monolayer graphene \cite{PRL103;226803}, silicene/Ag(111) \cite{PRB88;035432}, germanene/Au(111) \cite{SCR6;20714}, Si-adlayer/ZrB$_2$(0001) \cite{PRB90;075422}, and tinene/Bi$_2$Te$_3$ [Pb-adlayer/Au(111); \cite{PRB87;115138}], show the gapless $\&$ linear Dirac cone, the separated/modified one, the multi-Dirac-cone structures, their thorough absence, and the $p$-type semimetals near the Fermi level. The current viewpoints could be generalized to the other emergent 2D materials. For example, few-layer Ge-, Sn-, Pb-, Sb-, and Bi-related \cite{JPCC119;11896,SCR4;4794} systems are worthy of the systematic studies in the near future, since they possess more occupied electron orbitals and significant spin-orbital couplings. Such materials are under the current investigations.

\bigskip
\bigskip
\centerline {\textbf {ACKNOWLEDGMENT}}%
\bigskip
\bigskip

\noindent \textit{Acknowledgments.} This work was supported by the MOST of Taiwan, under Grant No. MOST 105-2112-M-022-001.

\newpage

\centerline {\Large \textbf {Table and Figure Captions}}


Figure 1: The optimal honeycomb lattices of single-layer group-IV systems, with the different buckled structures and lattice constants.

Table 1: The significant geometric parameters, the lattice constant, the height difference between A and B sublattices, the bond length, and the total ground state energy per unit cell for monolayer group-IV materials.

Figure 2: The geometric structures for sliding bilayer graphene/silicene systems under the specific path: (a) ${\delta\_{a},=0}$ [in unit of lattice constant; AA-bb], (b) ${\delta\_{a},=1/8}$, (c) ${\delta\_{a},=3/8}$, (d) ${\delta\_{a},=4/8}$. (e) ${\delta\_{a},=6/8}$, (f) ${\delta\_{a},=8/8}$ [AB-bt], (g) ${\delta\_{a},=11/8}$, (h) ${\delta\_{a},=12/8}$ along the armchair direction; (i) ${\delta\_{z},=1/8}$ $\&$ and (j) ${\delta\_{z},=3/8}$ along the zigzag one.

Table 2(a): The geometric properties of the sliding bilayer graphenes: lattice constant, interlayer distances $\Delta d$, Si-Si bond lengths $b$, and ground state energies per unit cell.

Table 2(b): The geometric properties of the sliding bilayer silicenes: lattice constant, interlayer distances $\Delta d$, Si-Si bond lengths $b$, and ground state energies per unit cell. In addition, the height differences between intralayer A and B sublattices $\Delta z$ and buckling angles $\theta$ are also shown.


Figure 3: Band structures of monolayer (a) C-, (b) Si-, (c) Ge-, (d)  Sn- and (e) Pb-related systems through the first-principles method, and those by the tight-binding model also shown in (f), (g) and (h) for C/Si/Ge [only the $\pi$ bonding],  Sn and Pb [the sp$^3$ bonding] , respectively.

Figure 4: The spatial charge distributions/their variations [${\rho}$/${\Delta\rho}$ for monolayer (a)/(b) carbon-, (c)/(d) silicon, (e)/(f) germanium-, (g)/(h) tin- and (i)/(j) lead-related systems with the ${[x, z]}$-plane projections. Those on ${[x, y]}$ plane are also shown in (a) and (b).

Figure 5: The four-orbital-projected density of states in monolayer (a) C-, (b) Si-, (c) Ge-, (d) Sn- and (e) Pb-generated systems.


Figure 6: Electronic energy spectra of sliding bilayer graphene systems for the stacking configurations: (a) ${\delta_a\,=0}$, (b) ${\delta_a\,=1/8}$, (c) ${\delta_a\,=4/8}$, (d) ${\delta_a\,=8/8}$, (e) ${\delta_a\,=11/8}$, (f) ${\delta_a\,=12/8}$ along the armchair direction; (g) ${\delta_z\,=1/8}$ $\&$ (h)${\delta_z\,=3/8}$ along the zigzag one.

Figure 7: The variations of spatial charge densities under the different stacking configurations in Figure 2, being illustrated on the ${[x, z]}$-planes.

Figure 8: The [2s, 2p$_x$, 2p$_y$, 2p$_z$]-decomposed van Hove singularities in sliding bilayer graphene materials arising from electronic energy spectra in Fig. 6.


Figure 9: Electronic energy spectra of sliding bilayer silicene systems for the stacking configurations: (b) ${\delta_a\,=0}$, (c) ${\delta_a\,=1/8}$, (d) ${\delta_a\,=3/8}$, (e) ${\delta_a\,=4/8}$, (f) ${\delta_a\,=6/8}$, (g) ${\delta_a\,=8/8}$, (i) ${\delta_a\,=11/8}$, (j) ${\delta_a\,=12/8}$ along the armchair direction; (k) ${\delta_z\,=1/8}$ $\&$ (l)${\delta_z\,=3/8}$ along the zigzag one. Also shown in (a)/(h) is that for monolayer silicene/AB-bt without the spin-dependent interactions.

Figure 10:  The spatial charge densities in (a)-(j) for sliding bilayer silicene systems with the same stacking configurations in Fig. 2, in which they are shown on the ${[x, z]}$-plane projection.

Figure 11: (a)-(l) The orbital-projected densities of states for sliding bilayer silicenes corresponding to band structures in Figs. 9(a)-3(l).

\newpage

\begin{figure}
\centering
\includegraphics[width=1\textwidth]{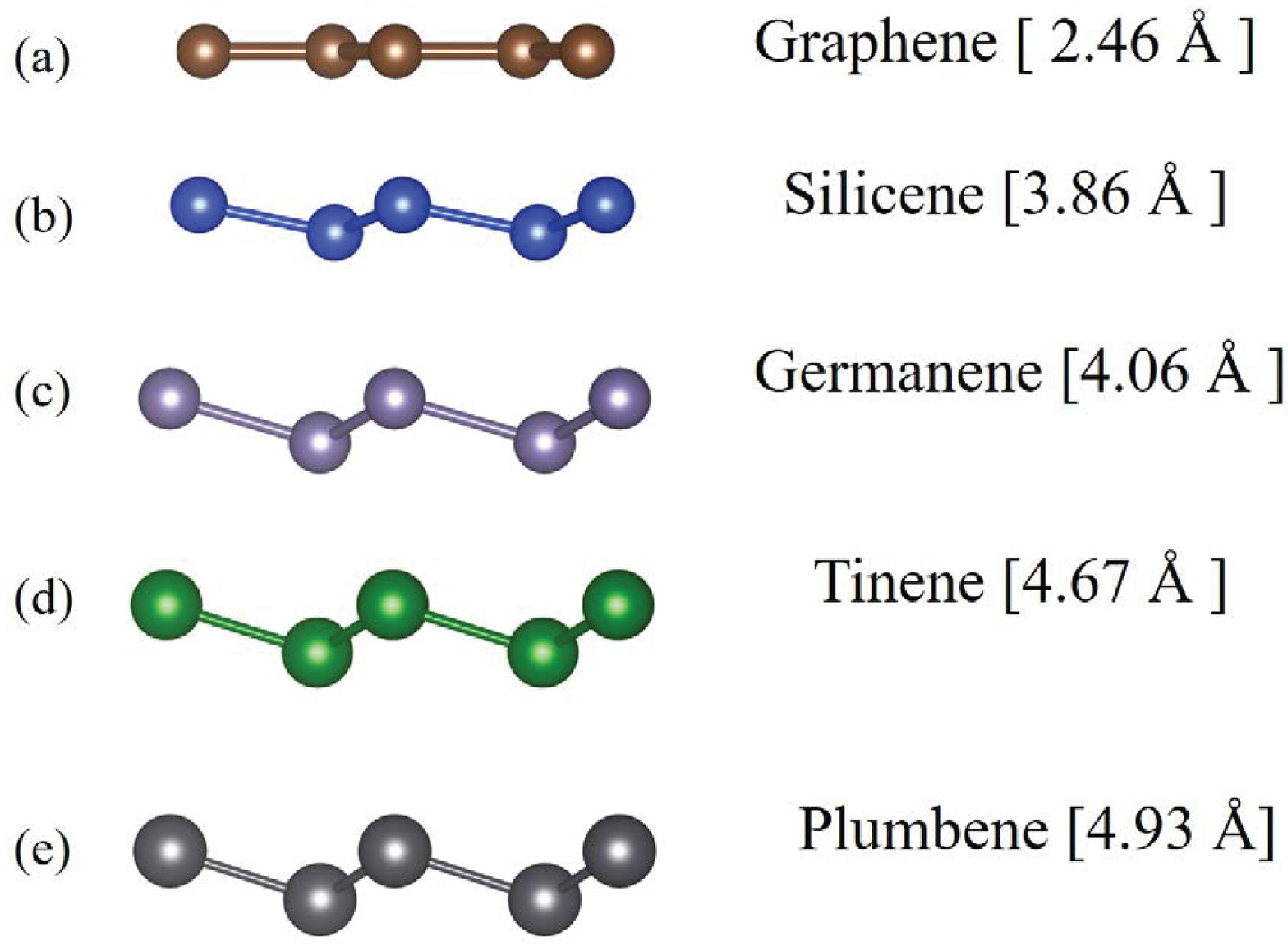}
\end{figure}

\newpage

\begin{figure}
\centering
\includegraphics[width=1\textwidth]{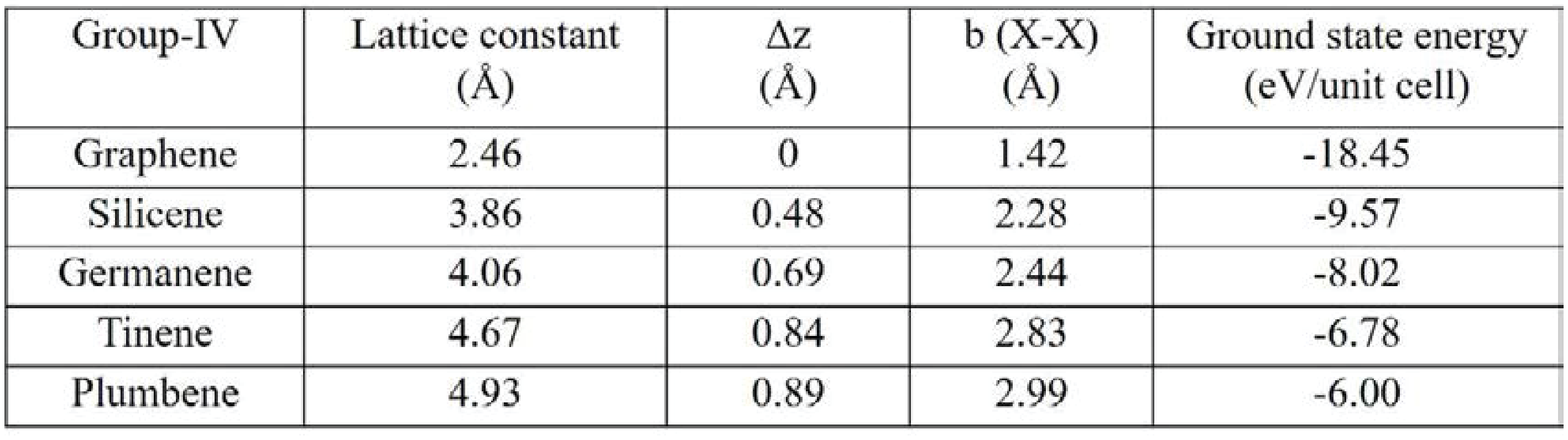}
\end{figure}

\newpage

\begin{figure}
\centering
\includegraphics[width=1\textwidth]{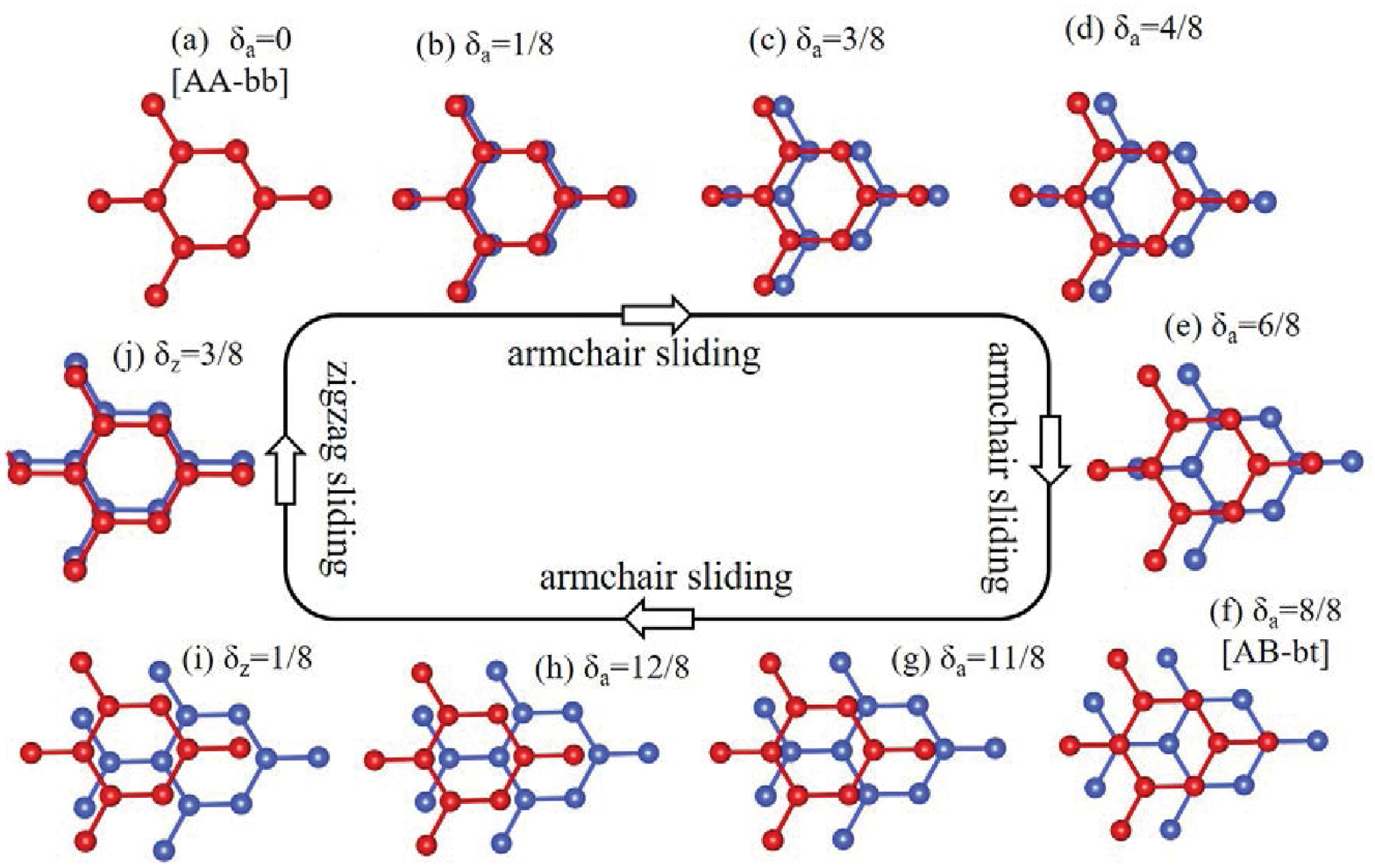}
\end{figure}

\newpage

\begin{figure}
\centering
\includegraphics[width=1\textwidth]{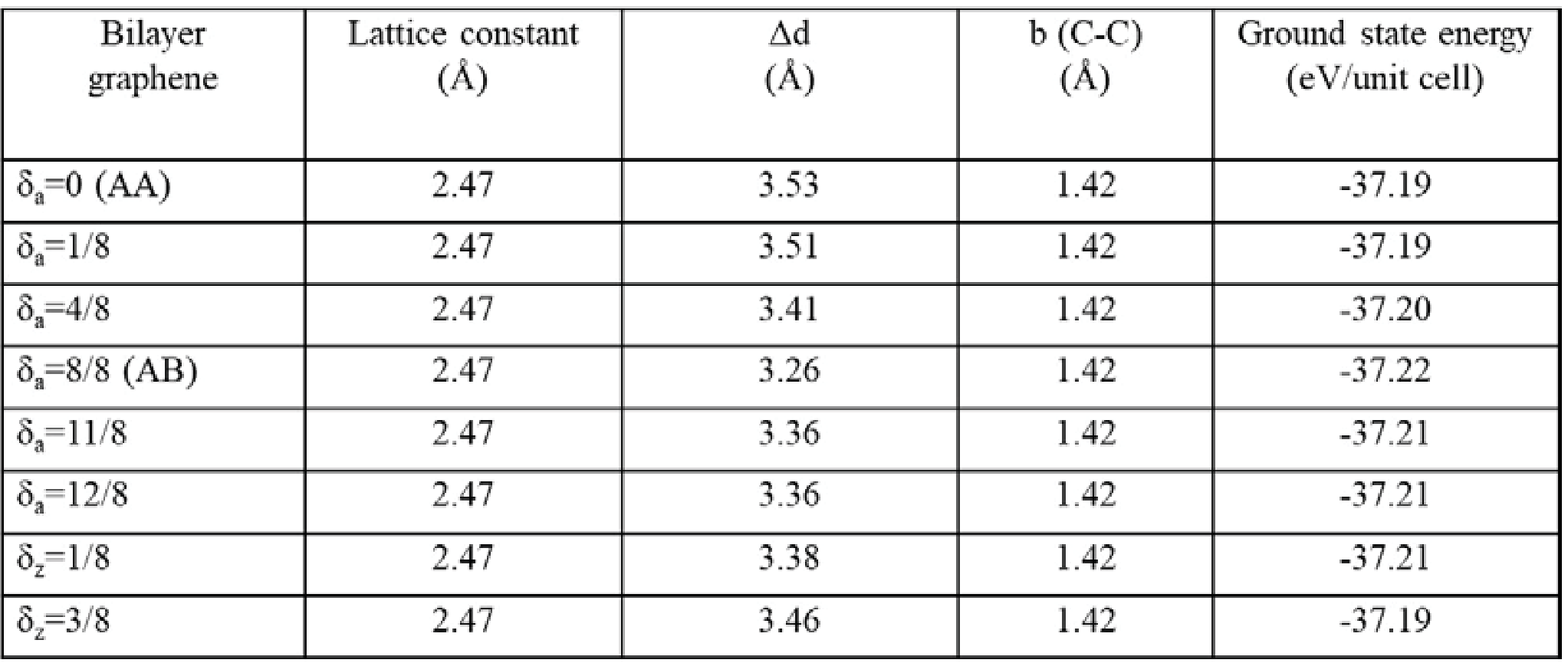}
\end{figure}

\newpage

\begin{figure}
\centering
\includegraphics[width=0.7\textwidth]{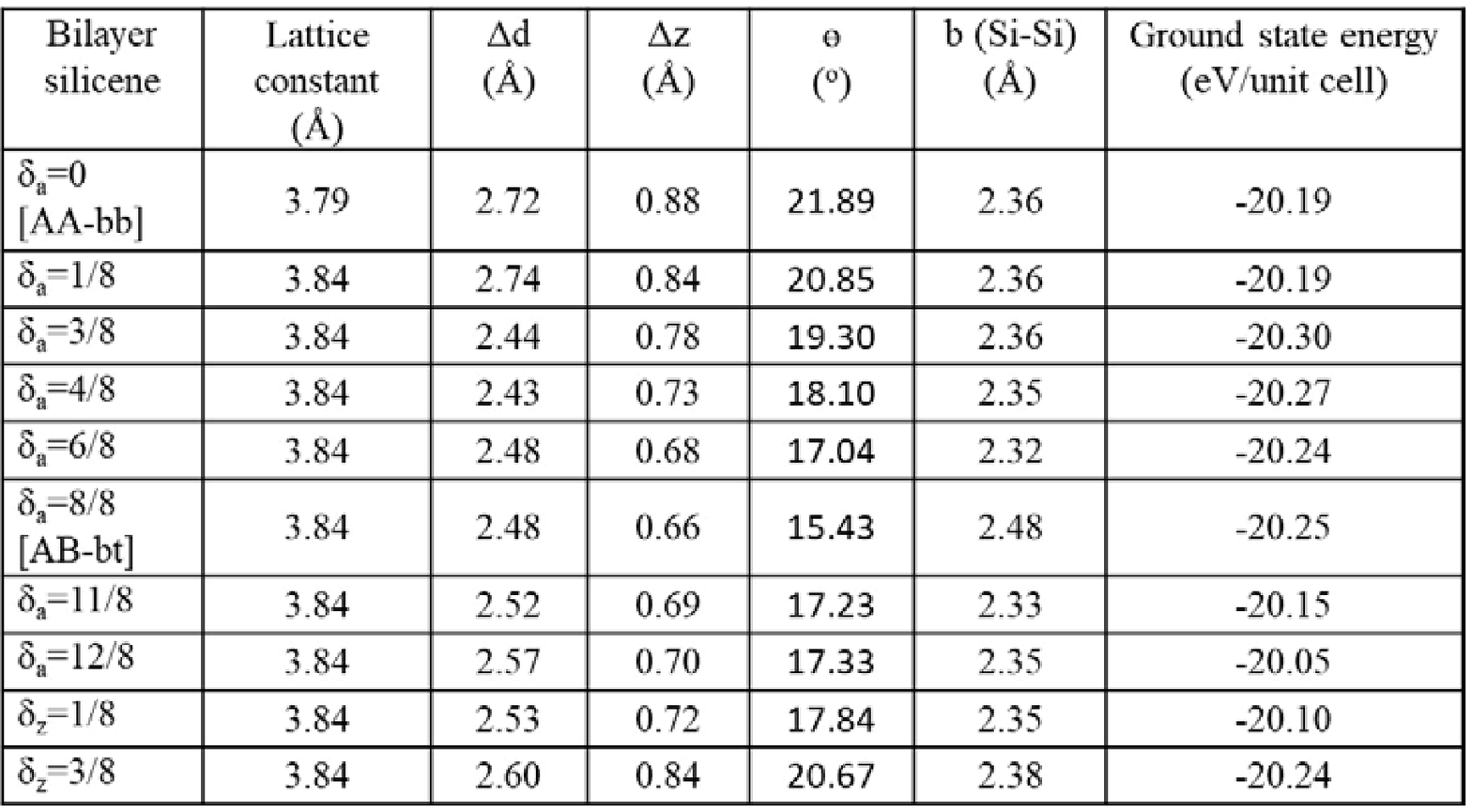}
\end{figure}

\newpage

\begin{figure}
\centering
\includegraphics[width=0.7\textwidth]{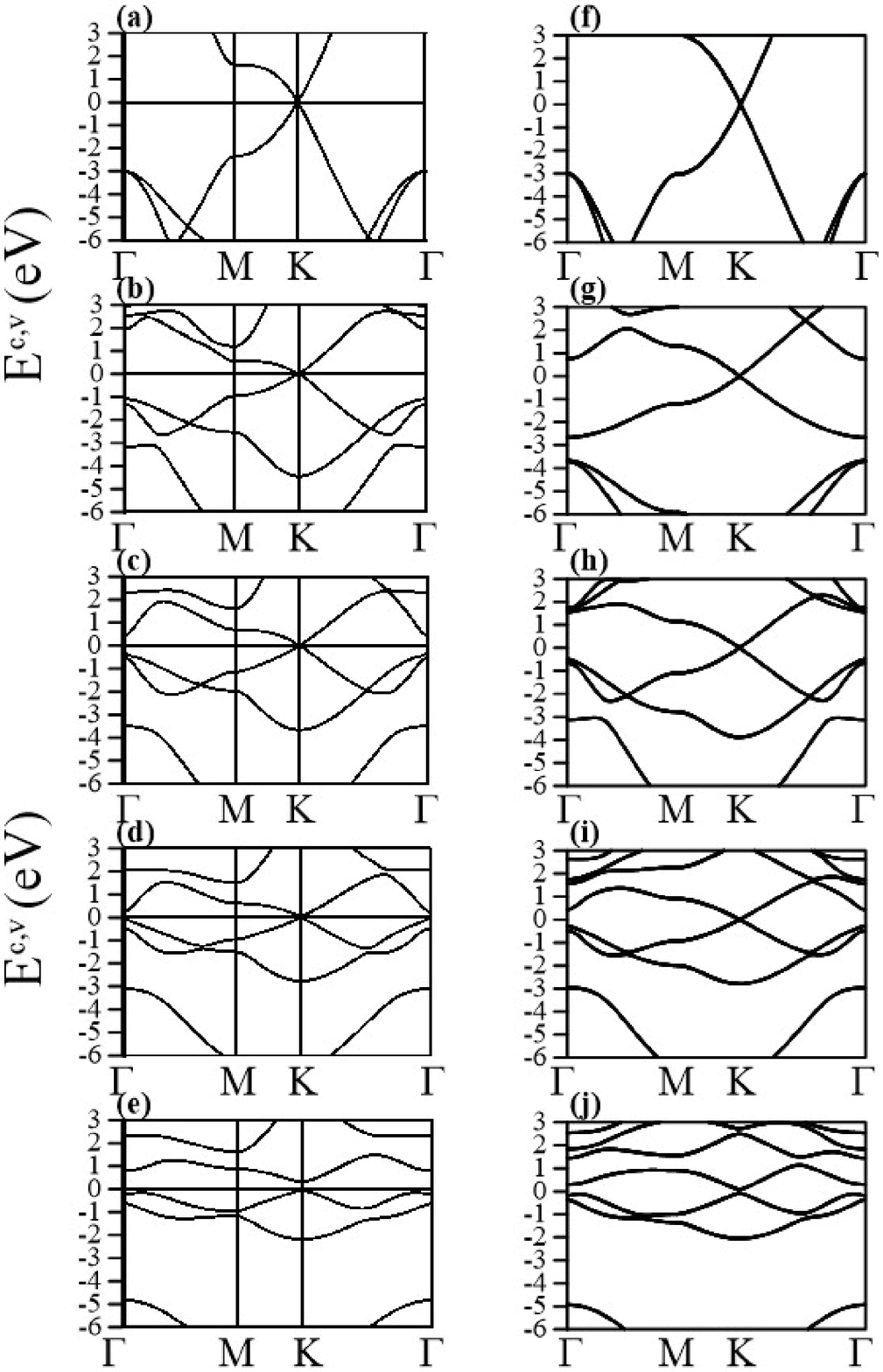}
\end{figure}

\newpage

\begin{figure}
\centering
\includegraphics[width=0.7\textwidth]{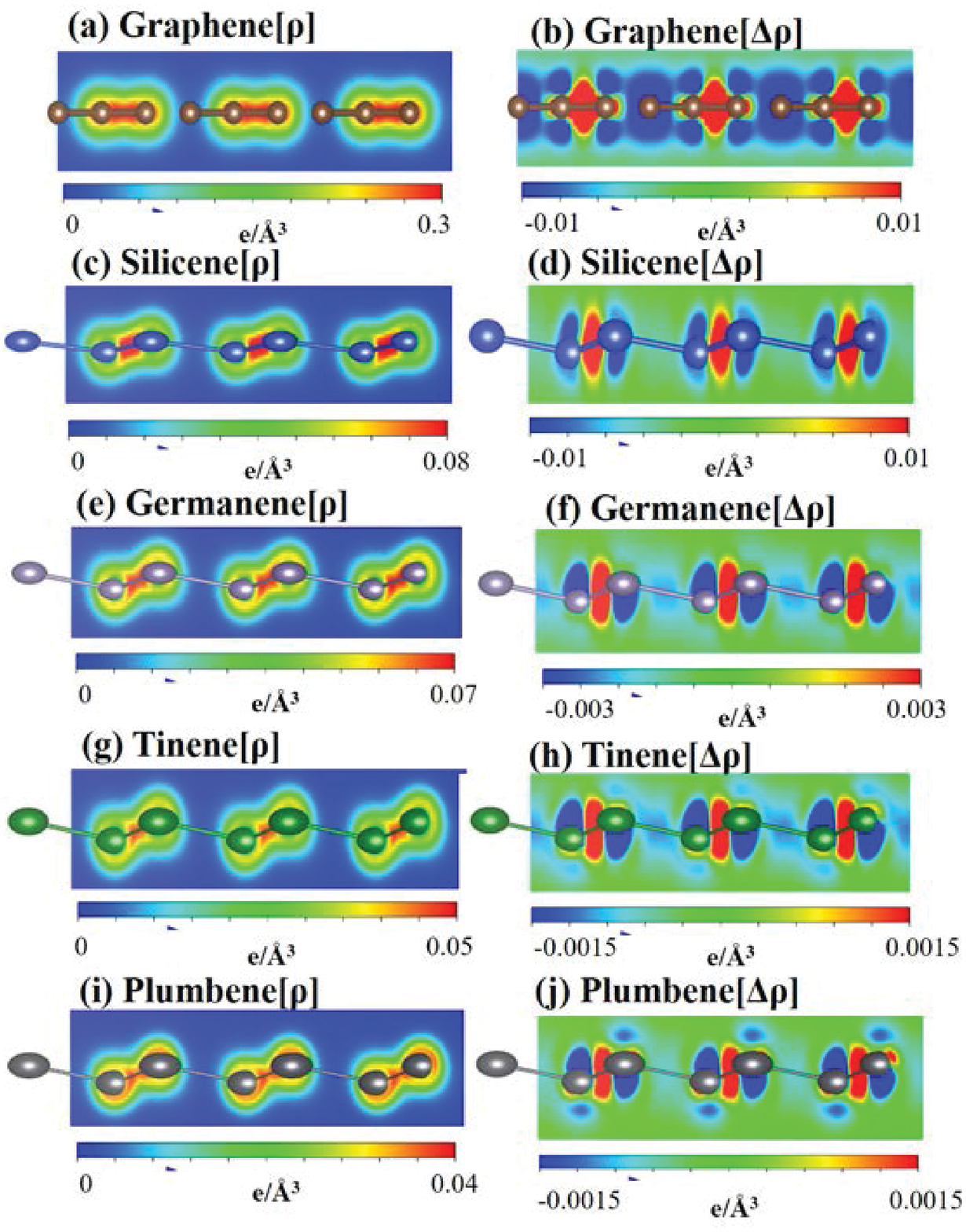}
\end{figure}

\newpage

\begin{figure}
\centering
\includegraphics[width=0.7\textwidth]{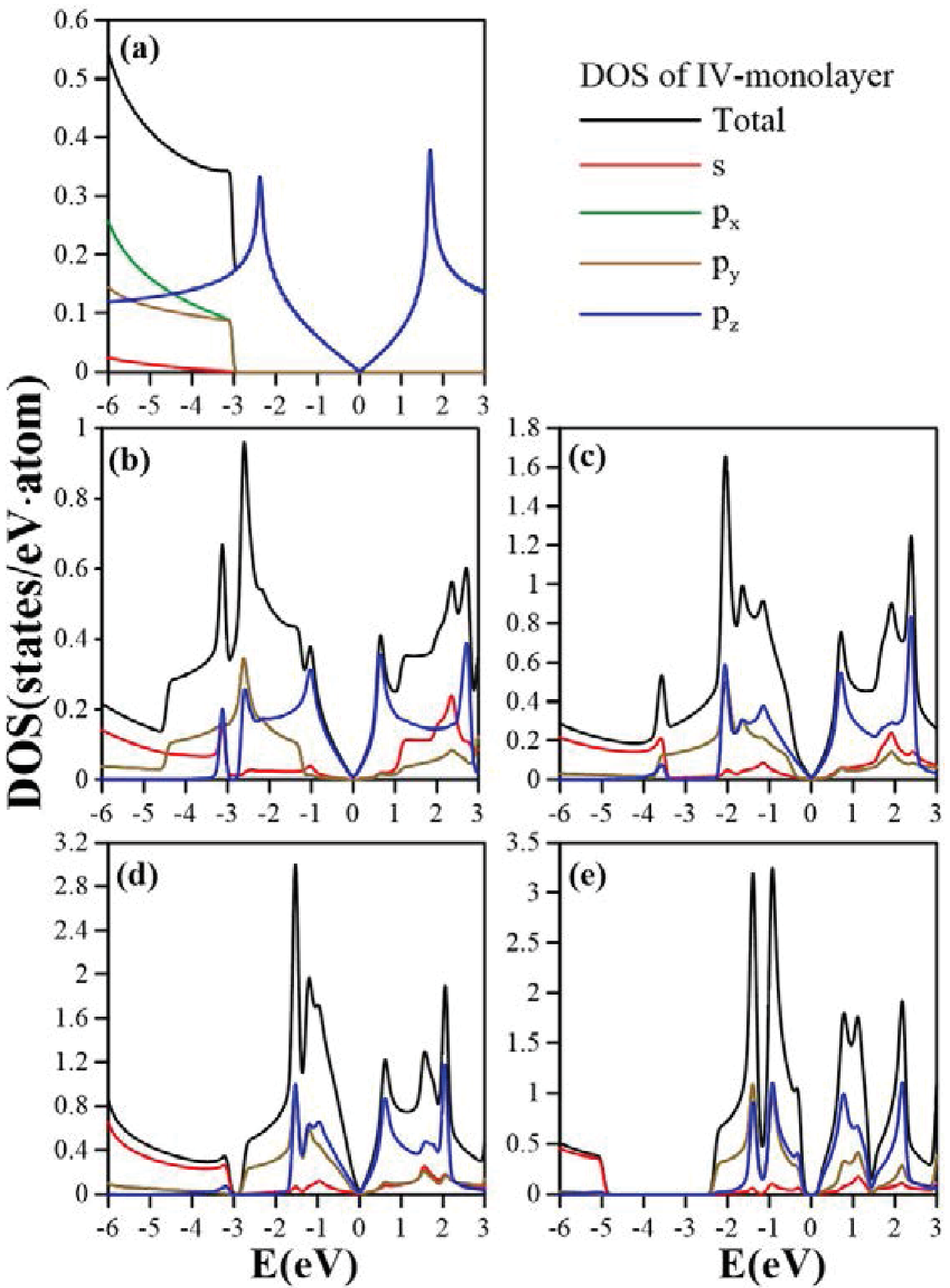}
\end{figure}

\newpage

\begin{figure}
\centering
\includegraphics[width=0.7\textwidth]{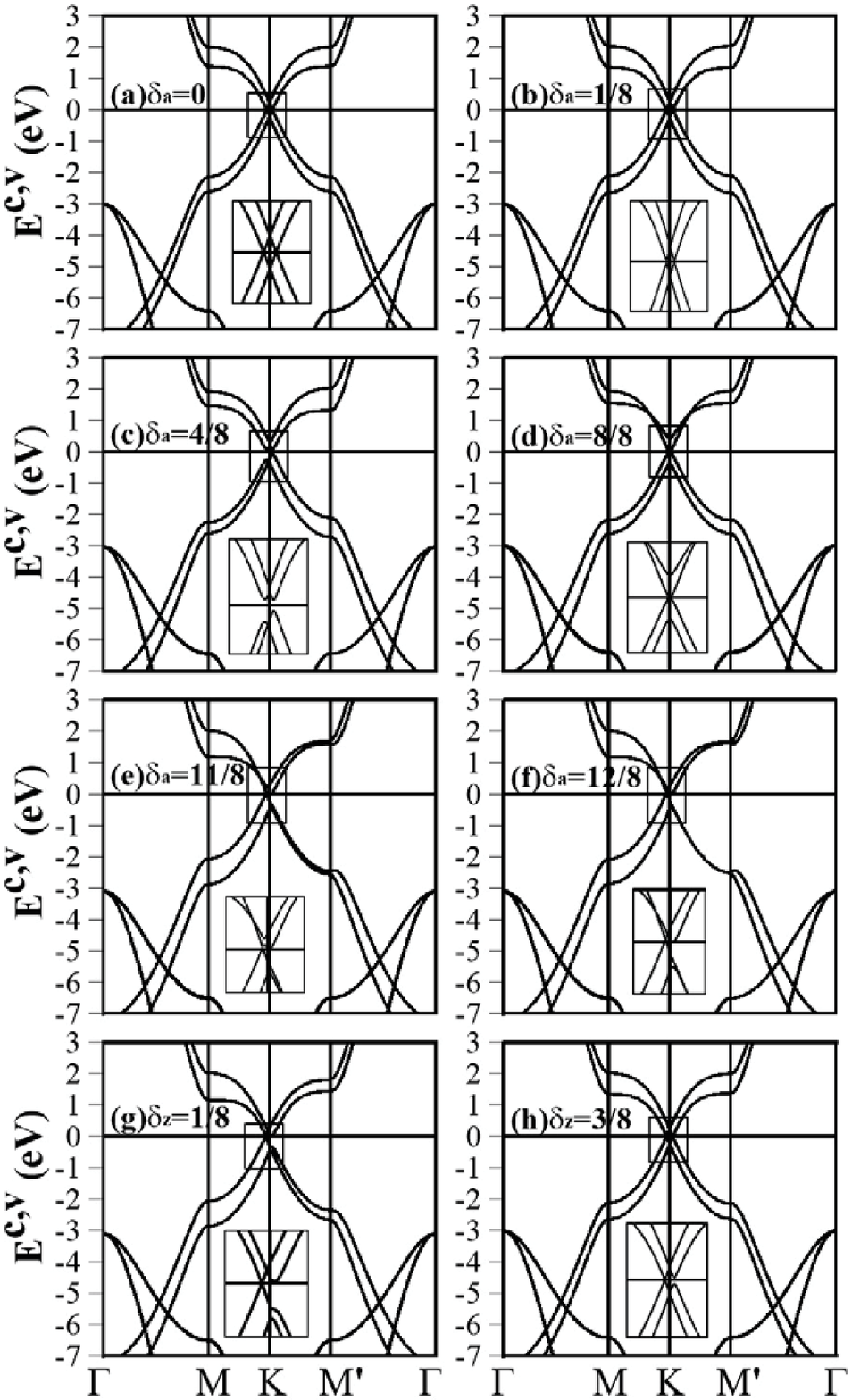}
\end{figure}

\newpage

\begin{figure}
\centering
\includegraphics[width=0.7\textwidth]{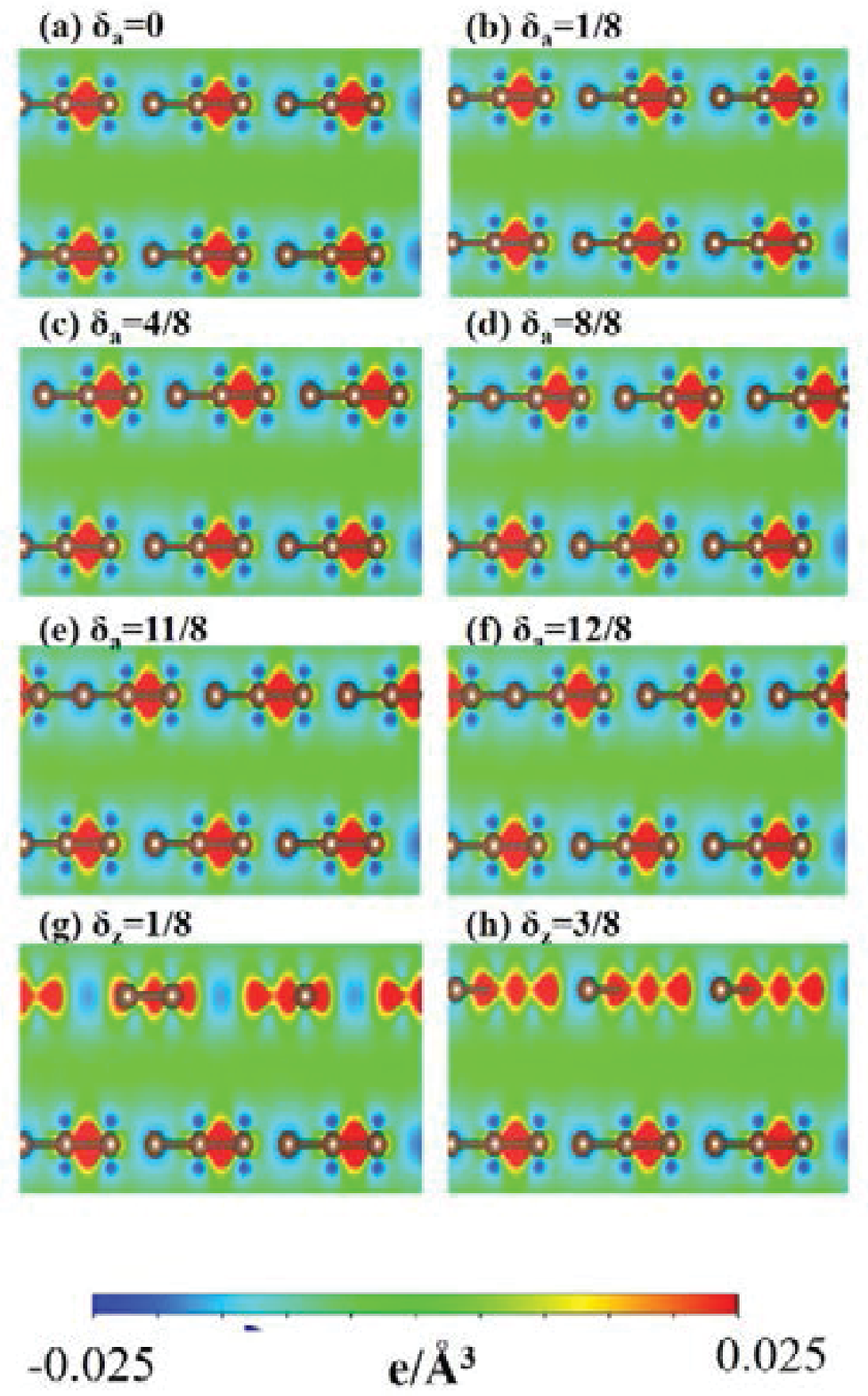}
\end{figure}

\newpage

\begin{figure}
\centering
\includegraphics[width=0.7\textwidth]{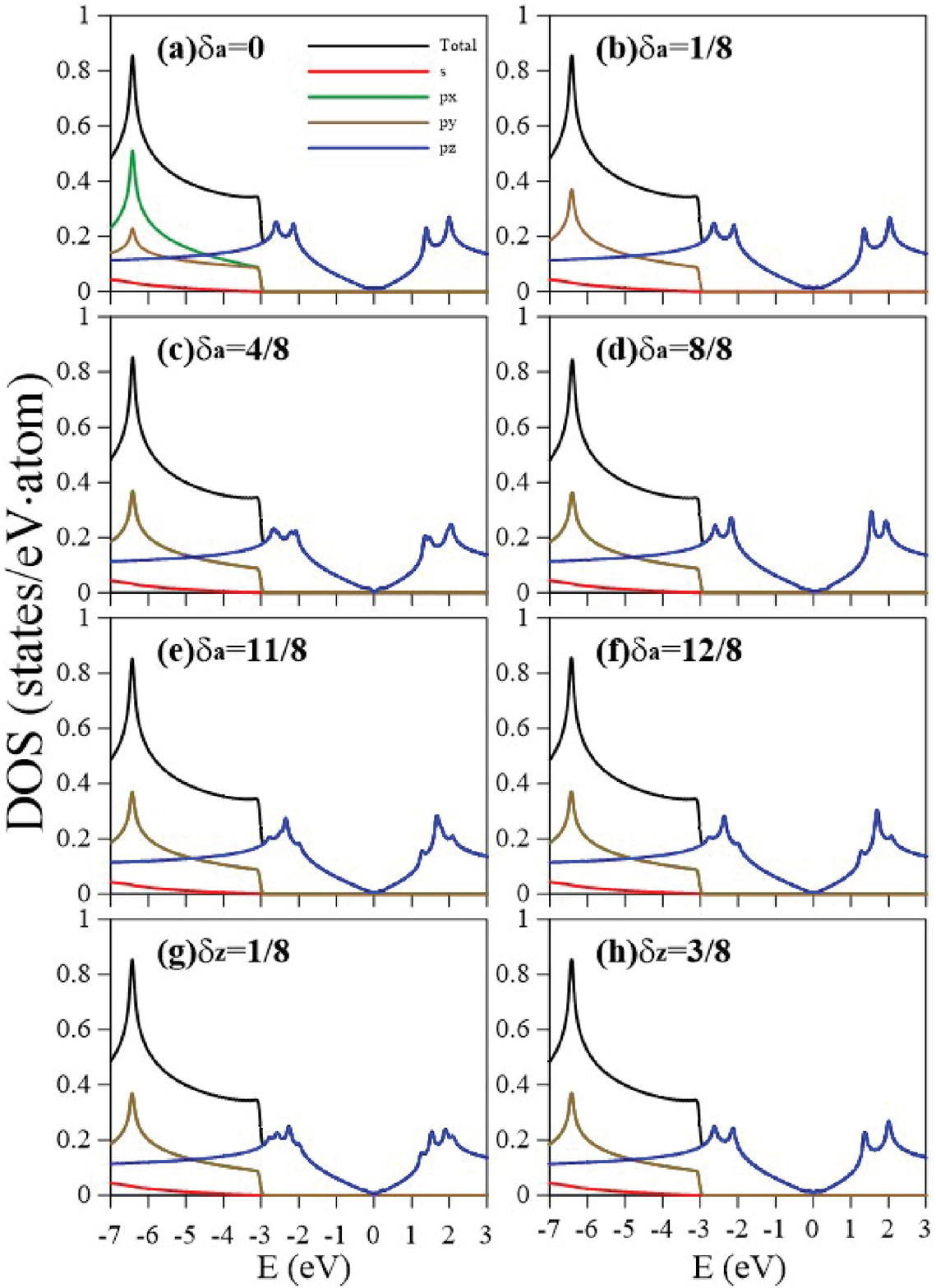}
\end{figure}

\newpage

\begin{figure}
\centering
\includegraphics[width=0.7\textwidth]{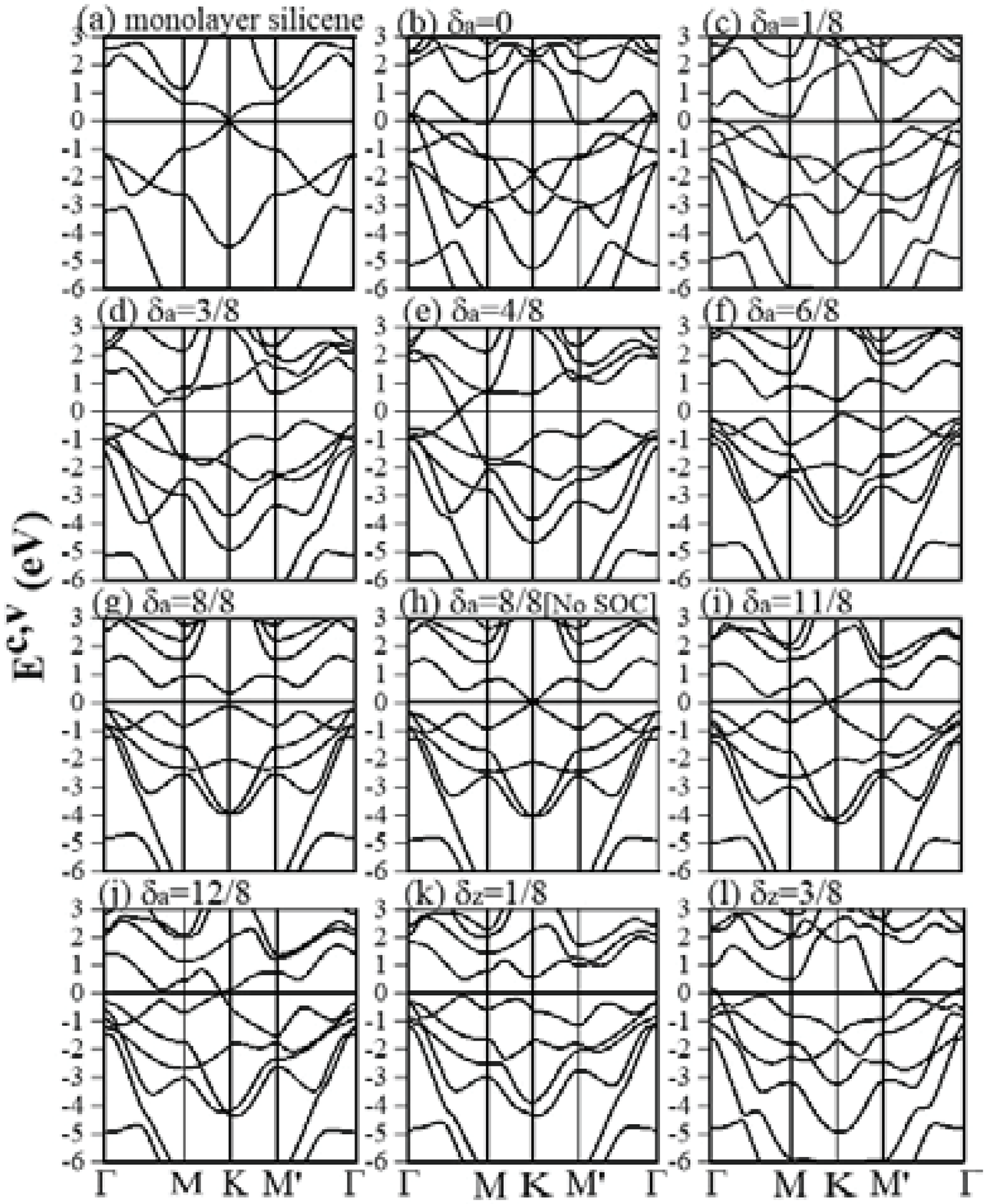}
\end{figure}

\newpage

\begin{figure}
\centering
\includegraphics[width=0.7\textwidth]{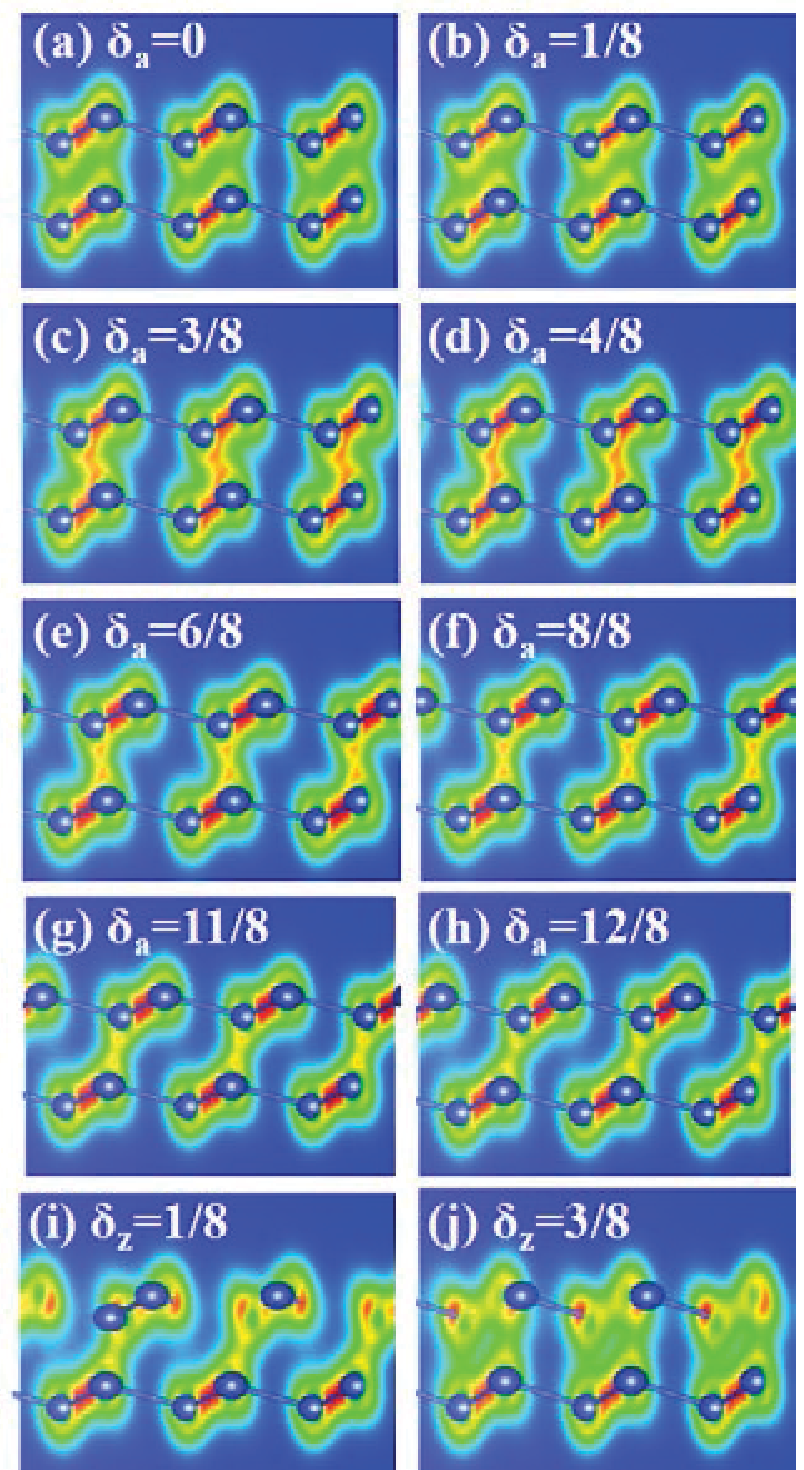}
\end{figure}

\newpage

\begin{figure}
\centering
\includegraphics[width=0.7\textwidth]{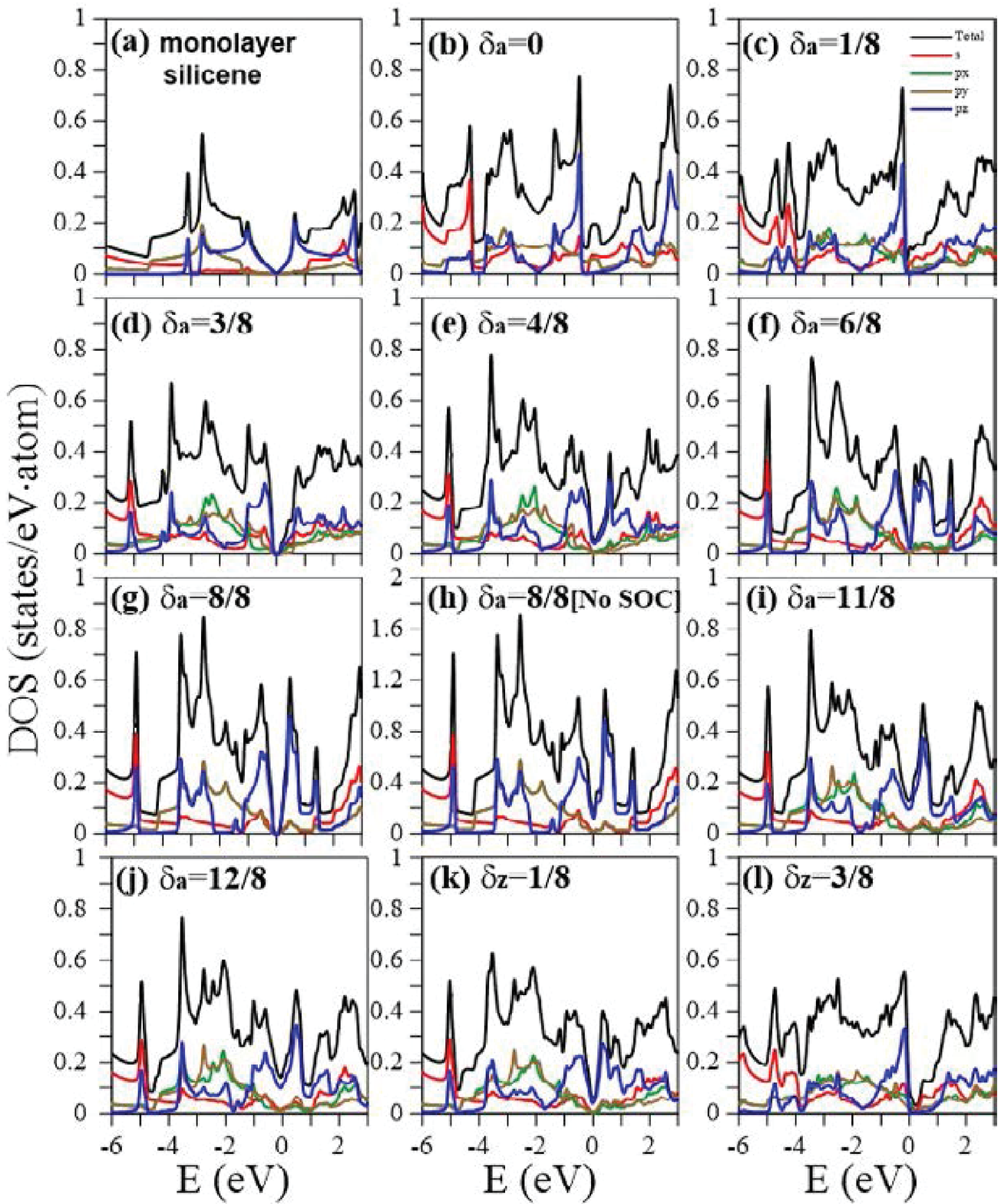}
\end{figure}

\end{document}